\documentclass[12pt]{iopart}

\usepackage{amsfonts}
\usepackage{graphicx}
\usepackage{yfonts}
\usepackage{dcolumn}
\usepackage{bm}

\usepackage{color}

\begin{document}


\title{Predictive power of MCT : \\ Numerical test and Finite size scaling for a mean field spin glass }

\author{Thomas Sarlat, Alain Billoire, Giulio Biroli}
\address{Institut de Physique Th{\'e}orique\\
CEA, IPhT, F-91191 Gif-sur-Yvette, France \\
CNRS, URA 2306, F-91191 Gif-sur-Yvette, France}

\author{Jean-Philippe Bouchaud}
\address{Science \& Finance, Capital Fund Management,\\
6, Bd Haussmann, F-75009 Paris, France}

\date{\today}

\begin{abstract}
The aim of this paper is to test numerically the predictions of the
Mode Coupling Theory (MCT) of the glass transition and study its
finite size scaling properties in a model with an exact MCT
transition, which we choose to be the fully connected Random
Orthogonal Model.  Surprisingly, some predictions are verified while
others seem clearly violated, with inconsistent values of some MCT
exponents. We show that this is due to strong pre-asymptotic effects
that disappear only in a surprisingly narrow region around the
critical point. Our study of Finite Size Scaling (FSS) show that
standard theory valid for pure systems fails because of strong sample
to sample fluctuations.  We propose a modified form of FSS that
accounts well for our results.  {\it En passant,} we also give new
theoretical insights about FSS in disordered systems above their upper
critical dimension.  Our conclusion is that the quantitative
predictions of MCT are exceedingly difficult to test even for models
for which MCT is exact. Our results highlight that some predictions
are more robust than others. This could provide useful guidance when
dealing with experimental data.
\end{abstract}

\pacs{64.70.Q,64.60.Ht,64.60.F}

\maketitle

\section{\label{sec:level1} Introduction }

In spite of all its shortcomings, the Mode-Coupling Theory 
(MCT)~\cite{MCTreviewdas,MCTbookGotze} of
the glass transition has most valuably contributed to our
understanding of the slowing down of super-cooled liquids.  It was the
first {\it ab initio} theory to make precise, quantitative predictions
about the appearance of a two-step relaxation process in super-cooled
liquids and hard sphere colloids as these systems become cooler or
denser.  MCT predicts in particular a non trivial
``$\beta$-relaxation'' regime where dynamical correlation functions
pause around a plateau value before finally relaxing to zero. Around
this plateau value, power-law regimes in time are anticipated,
together with the divergence of two distinct relaxation times,
$\tau_\alpha$ and $\tau_\beta$, as the MCT critical temperature $T_d$
is approached. Although this divergence is smeared out by activated
events in real liquids, the two-step relaxation picture suggested by
MCT seems to account quite well for the first few decades of the
increase of $\tau_\alpha$~\cite{MCTreviewdas,MCTbookGotze}.

Quite remarkably, it was observed soon after by Kirkpatrick,
Thirumalai and Wolynes~\cite{KIR1,KIR2}, that the MCT equations in
fact describe the {\it exact} evolution of the correlation function of
mean-field p-spin and Potts glasses~\cite{KIR3,reviewBCKM}. Furthermore, some
physical properties of these p-spin glasses (such as the existence of
an entropy crisis at a well defined temperature $T_s < T_d$) are
tantalizingly similar to those of super-cooled liquids, even beyond the
MCT regime. This has led to:
\begin{itemize}
\item A better understanding of the physical nature of the MCT
transition, in terms of a ``demixing'' of the unstable saddle points
of the energy landscape and the (meta)stable minima: above a certain
energy threshold $E_{th}$ only the former type is found, while minima
all sit below this threshold (for a recent review, see ~\cite{Cavagnareview}).
\item The elaboration of a consistent phenomenological description of
the glass transition, called the ``Random First Order Transition'' (or
RFOT) theory by Wolynes and 
collaborators~\cite{KIR1,KIR2,Lubchenko:yq}, within which the activated
processes allowing the system to hop between metastable minima is
given a precise interpretation in terms of spatial rearrangements.
\end{itemize}

More recently, it has been argued that MCT can be thought of as a
Landau theory of the glass transition, where the order parameter is
the (small) time dependent difference between the correlation function
and its plateau value~\cite{ABB}. It was furthermore shown that the
MCT transition is accompanied by the divergence of a dynamical
correlation length~\cite{BB-EPL,IMCT} (see also \cite{chi4a}), which gives a quantitative
meaning (within MCT) to the concept of heterogeneous dynamics. Correspondingly,
critical fluctuations are expected close to the MCT transition $T_d$,
and become dominant in spatial dimensions $d < d_u$, where the value
of the upper critical dimension is $d_u=6$ or $8$ depending on the
existence of conserved variables (see~\cite{BB-SE,BBBKKR}). For the
model considered below, $d_u=6$.

In view of the central role of the Mode-Coupling Theory in our current
understanding of the glass transition, it is somewhat surprising that
so little numerical work has been devoted to models for which the MCT
equations are believed to be exact.  To our knowledge there is in fact
no exhaustive treatment of a spin model displaying an exact
MCT-transition. Such a study is valuable for several reasons. First,
it is important to know how precisely the MCT predictions can be
tested on a model where the theory is supposed to be exact, and for
which all the excuses for MCT's failures in real systems (uncontrolled
approximations, activated events, low dimensions, etc.) are
absent. Second, the detailed study of finite size effects is
important, since one expects in that case to observe in a controlled
way the famous crossover between the MCT regime and the activated
regime. Furthermore, since the short-range nature of the interactions in liquids
should somehow lead to finite size corrections to the MCT equations,
the results of this analysis should bring important insights, and perhaps help to
understand the somewhat unexpected results of Karmakar et al. \cite{IndiansPNAS}. 

One reason explaining why these numerical studies are scarce is the
slow dynamics of these models. Even more so than in other spin glass
models a good sampling requires a
number of Monte Carlo iterations that grows rapidly with the system
size even with an extremely efficient sampling algorithm, like the
parallel tempering algorithm \cite{Marinari:1992qy}. An other reason is that the paradigmatic
p-spin model is a p-body interaction model, which only leads to
MCT-like dynamics for $p \ge 3$. Altogether this leads to quite heavy
simulations (see~\cite{BIL}). Other mean field models in the MCT class
exist, like the fully-connected q-states Potts model (for $q\ge 4$) \cite{KIR1} or
the random orthogonal model (ROM)~\cite{PAR2,ParisiPotters}, which are two-body spin
models. But a recent detailed study of the 10-states Potts
model~\cite{BRA1,BRA2,BRA3,Brangian:uq} has led to rather strange
results: absence of the ``cage effect'', unusual finite-size scaling
exponent. Nevertheless some behavior reminiscent of the MCT-transition
seems to emerge. The ROM case is even worse; there is no precise study
in the temperature range close to the dynamic temperature, where all
previous simulations have fallen out-of-equilibrium even for rather
small system ($N = 50$~\cite{PAR2,RIT2,RIT3,LEF1}).

Our project is to provide an exhaustive numerical analysis of the
statics and the dynamics of a model for which the Mode Coupling Theory
is exact. We want in particular to : (i) test numerically the MCT
predictions concerning the two-point relaxation function ($q_d(t)$)
and the four-point correlation functions ($\chi_4(t)$) that describe
dynamical heterogeneities, (ii) study the pre-asymptotic corrections
to the critical behavior and (iii) analyze the finite size scaling of the MCT 
transition. \\
To achieve such a program, one needs a
two-body model with well-separated static ($T_s$) and dynamic ($T_d$)
transition temperatures.  We have found that a certain variant of the
ROM satisfies these constraints.  We also need an efficient algorithm,
since the ROM turns out to be an extremely difficult model to
simulate. If one uses a simple algorithm like thermal annealing, the
convergence is so poor that the average energy never goes below
$E_{th}$ below $T_d$, even for small system sizes.  In this work we
will use the best algorithm to date for spin glass numerical
simulations, namely the {\it parallel tempering} algorithm \cite{Marinari:1992qy}. In spite
of tremendous numerical efforts, we are still limited to rather small
systems (up to $256$ spins). However, our simulations allow us to
reach an interesting but somewhat unexpected conclusion: preasymptotic
corrections and finite size effects are so strong that a direct
observation of the MCT predictions is extremely difficult. Great care
must be exercised to extract meaningful value of critical exponents,
even for a model for which MCT is in principle exact! Such problems
are expected to arise in the analysis of experimental data as well,
and we believe that our work highlights some caveats concerning the
validity and relevance of MCT in practice.

Our first main result consists in establishing that pre-asymptotic effects are extremely important
until one approaches the MCT transition temperature up to fractions of percent!
We show this effect in two ways: (a) our numerical simulations of the ROM unveil
that  several of the MCT predictions are still not verified quantitatively for temperatures a few percent higher than the critical one; 
(b) we solve numerically the schematic Mode Coupling Theory equations and show that very 
similar conclusions are also reached in this framework. 
Similar findings were obtained in \cite{albakrakoviacJCP2002} by comparing MCT to experimental data. 

Our second main result is a detailed study of finite size scaling of the MCT transition.  
This is both of practical and theoretical interest when one attempts to determine some MCT exponents. 
We find that sample to sample fluctuations play a crucial role. 
As we shall explain, thermal fluctuations would lead naturally to finite size effects
that becomes relevant for distance from the critical temperature of the order 
$N^{-2/3}$ where $N$ denotes the system size.  
However, disorder fluctuations, that is sample to sample fluctuations of the
dynamic temperature, are of order $N^{-{1}/{2}}$. The latter
fluctuations therefore dominate and lead to a very subtle finite size scaling
behavior. Our study lead us to a generalization of the Harris criterion suitable for 
disordered systems above their upper critical dimension.

The paper is organized as follows. After recalling the main predictions of the replica method (for the
statics) and of the MCT (for the dynamics) in section~\ref{sec:2}, we
discuss our numerical results for the statics of the ROM in
section~\ref{sec:3}. In~\ref{sec:dyn}, we investigate the equilibrium
dynamics of the model and compare the results with the predictions of
MCT. We show that the glaring discrepancies come from the unexpectedly
narrow critical window around the MCT transition point, which we
cannot access numerically without large finite size effects.  We then
turn in section~\ref{sec:4} to a detailed study of these finite size
effects, and to the possibility of using dynamical finite-size scaling
for this model. We find that naive finite-size scaling theory fails to
account for our results, because of the strong sample to sample
dependence of the critical temperature. We show how to understand
phenomenologically our results in the rest of section~\ref{sec:4},
relegating to appendices more precise statements on the apparent
breakdown of the Harris criterion in high dimensions, and the exact
solution of the fully-connected disordered Blume-Capel model, which
provides an explicit illustration of our arguments. The last section
is dedicated to the conclusion and open questions.

\section{\label{sec:2} 1-RSB models: A short summary of known results}

As mentioned in the introduction, the starting point of the Random
First Order Theory of the glass transition is the analysis of 
the so-called discontinuous spin glasses, namely
mean field disordered models with an
order parameter that has a  jump at the transition. In terms of the replica
method, these are the ones solved by the so-called one-step replica
symmetry breaking ansatz (1-RSB, see~\cite{Cugliandolo:ao,Cavagnapedestrians} and
references therein).  Examples are the p-spin (spherical) model and the
Random Orthogonal Model that we are going to study in detail in the
following.

The physical behavior of these models is particularly transparent in
terms of the Thouless-Anderson-Palmer (TAP) \cite{TAP} approach that allows to
analyze the free energy landscape of the model. Technically, this is a
Legendre transform of the free energy as a function of all the local
magnetizations. The minima of the TAP free energy correspond to the
thermodynamic states of the system, like the two minima of the
Curie-Weiss free energy represent the two low temperature
ferromagnetic states in the (completely connected) Ising model.

For a 1-RSB system, the analysis of the number $\mathcal{N}$ of solutions
of the TAP equations that contribute to the free energy density shows
that there exists two transition temperatures: one static ($T_s$) and
the second dynamic ($T_d$), with $T_d>T_s$.  When $T>T_d$ or $T<T_s$,
the complexity, defined as $N^{-1} \ln \mathcal{N}$, vanishes in the
large $N$ limit ($N$ is the total number of spins). Above $T_d$, this
is because the energy is higher than the threshold value $E_{th}$
mentioned in the introduction, so that typical states are unstable
saddles, hence the system is in a paramagnetic state.  Below $T_s$, this is
because the low lying amorphous minima are not numerous
enough. Actually, the thermodynamic transition is precisely due to the
vanishing of the complexity at $T_s$. Below $T_s$ the system is in the
so-called 1-RSB phase (correspondingly the high temperature phase is
called replica symmetric (RS)). For $T_s<T<T_d$, on the other hand,
the complexity takes a finite nonzero value. In other words, between
$T_s$ and $T_d$, an exponential number of metastable states
contributes to the free energy. In this temperature range, the system
is already not ergodic since a configuration starting in one of these
multiple states and evolving with say, Langevin dynamics remains
trapped inside the initial state forever (i.e. on all timescales not
diverging with the system size).

When approaching $T_d$ from above, the majority of the stationary
points of the TAP free energy, which are unstable above $T_d$, become
marginally stable at $T_d$ and stable below. One therefore expects a
slowing down of the dynamics due to the rarefaction of descending
directions in the free energy landscape~\cite{lalouxkurchan}. In fact,
the dynamics of some of these models, e.g. the p-spin spherical model,
can be analyzed exactly. One can show, from the dynamical equation for
the spin-spin time-dependent equilibrium correlation functions, that
at $T_d$ an infinite plateau appears in the dynamical overlap $q_d(t)$
defined as:
\begin{eqnarray}
q_d(t)=\frac{1}{N} \sum_{i} \langle  \sigma_i(0)\sigma_i(t) \rangle,
\label{qddt}
\end{eqnarray}
where $\sigma_i$ is the spin at site $i$.  Remarkably, the
integro-differential equations derived with the Mode Coupling Theory
of Glasses for the density-density correlation functions reduce
identically within the so-called schematic
approximation~\cite{gotzerev} to the equation obeyed by $q_d(t)$.  The
solution of the equation for $q_d(t)$ (or the more complicated full
MCT equations)~\cite{CHA} leads to a two-step relaxation for the
correlation function: there is a first rapid decay from 1 towards a
plateau value $q_{EA}$, then a slow evolution around it (the $\beta$
regime) and, eventually, a very slow decay from it (the $\alpha$
relaxation).  The plateau value is called the Edwards-Anderson
parameter, in analogy with spin glasses.  For our present purpose, we
will be interested in these two last regimes.

At a given temperature $T>T_d$, the dynamics in the $\beta$-regime is
described by power-laws. The approach to the plateau can be
written as $q_d(t) \sim q_{EA}+ct^{-a}$, and the later departure from the
plateau as $q_d(t)\sim q_{EA}-c't^{b}$. The exponents $a$ and $b$ are model
dependent, but satisfy the  universal relation~\cite{gotzehouches,ABB}:
\begin{eqnarray}
\frac{\Gamma^2(1+b)}{\Gamma(1+2b)}=\frac{\Gamma^2(1-a)}{\Gamma(1-2a)}\ .
\label{eq:exp}
\end{eqnarray}
where $\Gamma(x)=\int_0^\infty t^{x-1}e^{-t}dt$ is the Gamma function.
In the $\alpha$-regime, close to $T_d$, $q_d(t)$ verifies a scaling
law, called time-temperature superposition in the structural glass
literature:
\begin{equation}
q_d(t) \simeq f(t/\tau_\alpha)\ .
\label{tts}
\end{equation} 
This scaling form allows one to superpose on a single curve the data
for different values of time and temperature. A good fit of the
function $f(x)$ is obtained with  a stretched exponential
$f(x)\propto \exp(-x^{\beta})$. The $\alpha$ relaxation time
$\tau_\alpha$ diverges at the transition $T=T_d$, as:
\begin{eqnarray}\label{eq:gamma}
\tau_\alpha &\propto& (T-T_d)^{-\gamma} \ , \nonumber \\
\gamma&=&\frac{1}{2a} + \frac{1}{2b} \ .
\label{eq:tau}
\end{eqnarray}

The dynamical correlation $q_d(t)$ is the order parameter of the
dynamical MCT transition.  Recently, especially in connection with the
phenomenon of dynamical heterogeneity in
glass-formers~\cite{edigerreview}, there has been a lot of interest in
the critical fluctuations and dynamical correlations associated with
this transition. A central observable introduced
in~\cite{dasguptaetal,parisifranzdonatiglotzer} is the four-point
susceptibility $\chi_4(t)$, which measures the thermal fluctuations of
the order parameter $q_d(t)$:
\begin{equation}\label{eq:defchi4}
\chi_4(t)=N(\overline{\langle q_d(t)^2 \rangle - \langle q_d(t) \rangle^2})\,
\end{equation}
where, as usual, the brackets denote a thermal average and the
over-line denotes an average over the quenched random couplings.  For
mean field glass models, or more generally for MCT (without conservation laws), one can show that
the four-point correlation function becomes critical near the MCT
transition temperature $T_d$~\cite{toninelli:041505,IMCT}. In the
$\beta$-regime, and with $\epsilon=(T-T_d)/T_d$,
\begin{eqnarray}
\chi_4(t) \simeq \frac{1}{\sqrt{\epsilon}}\,
f_1(t\epsilon^{\frac{1}{2a}}) \ , \quad t \sim
\tau_\beta=\epsilon^{-\frac{1}{2a}}
\end{eqnarray}
and in the $\alpha$-regime,
\begin{eqnarray}
\chi_4(t) \simeq \frac{1}{\epsilon}\, f_2(t\epsilon^{\gamma}) \ , \ t\sim \tau_\alpha \ ,
\end{eqnarray}
where $f_1(x)$ and $f_2(x)$ are two scaling functions, with the
following properties: $f_1(x) \propto x^a$ when $x \ll 1$ and scales like $x^b$
when $x\gg1$; $f_2(x)$ scales like $x^b$ for $x\ll 1$ and vanishes for
large $x$ (note that the large time limit of $\chi_4(t)$ is equal to the
spin-glass susceptibility which is  not critical at
$T_d$). Using the definition of $\gamma$, and equation (\ref{eq:exp}),
the numerical analysis of $\chi_4(t)$ very close to $T_d$ allows,
at least in principle, to extract all the MCT-exponents without having
to know the value of $q_{EA}$~\cite{toninelli:041505}.

Finally, in~\cite{BB-EPL,IMCT} it has been shown that MCT can be
interpreted as a mean field approximation  of a critical model. Following
this point of view, one can compute, within this mean field theory,
the spatial dynamical correlations and the associated diverging
correlation length at the transition $\xi(T)$. It can be established
that $\xi(T)\propto 1/(T-T_d)^\nu$ with $\nu=1/4$ (see also \cite{FranzMontanari}). Furthermore, the
upper critical dimension of the theory turns out to be $d_{u}=6$ for
dynamics without exactly conserved variables~\cite{BB-SE,BBBKKR}.

\section{\label{sec:3} Numerical 
simulations of the Random Orthogonal Model: equilibration and static
properties}

\subsection{Definition of the model}

The ROM~\cite{PAR2} is a fully-connected spin model with quenched
disorder, defined by the Hamiltonian:
\begin{eqnarray}
H=-\frac{1}{2} \sum_{ij} \sigma_{i} J_{ij} \sigma_{j},\qquad J=^{t}O\Lambda O\ ,  \nonumber
\end{eqnarray}
where $\Lambda$ is a diagonal matrix whose elements are equal to $\pm 1$ and are drawn
according to:
\begin{eqnarray}
\rho(\lambda)= p \delta(\lambda-1) + (1-p) \delta(\lambda+1)
\ ,
\end{eqnarray}
and $O$ is an orthogonal matrix distributed with respect to the Haar
measure (that we generated numerically by using 
the NAG routine G05QAF), $p$ is a real number $p\in [0,1]$. 
The will use later the notation $\omega\equiv\{J_{i,j}\}$ to
denote a given instance of the disorder.
The original model
corresponds to $p=1/2$. The normalisation is such that $\overline{J_{ij}^2}=1/N^2 \Tr J^2 =  1/N$, $\forall p$.  A
detailed analysis for arbitrary $p$ can be found in~\cite{LEF1}.  Both
the static and the dynamic temperatures depend on $p$ (see~\cite{LEF1}
for more details). 
In the following, $p$ is set to
$13/32\simeq 0.4$. This gives us higher transition temperatures than
for $p=1/2$ with a good separation of $T_s$ and $T_d$: $T_s \simeq
0.102 $ and $T_d \simeq 0.177$ respectively.\footnote{There might be other another
transition to a full RSB state at lower temperatures, but we will not be concerned by this
possibility.} With this value of $p$,
the annealed entropy vanishes at a temperature  close to $T_s$
(For $p>1/2$ the annealed entropy vanishes exactly at $T_s$).

The static order parameter of the 1-RSB transition is the usual static overlap $q$ between two
replicas, i.e.  two equilibrium configurations $\{\sigma\}$ and
$\{\tau\}$ characterized by the same quenched disorder. Its
probability distribution function can be written as:
\begin{eqnarray}
P(q)\equiv \overline{\langle
\delta(q-\frac{1}{N}\sum_i{\sigma_i}{\tau_i}) \rangle}.
\end{eqnarray}
In the thermodynamic limit, $P(q)=\delta(q)$ for $T>T_s$, two
different replicas have zero overlap with probability one. In the low
temperature (1-RSB) phase $T<T_s$, a second delta function peak
centered at a value $q_1>0$ appears with a weight $1-m$. To wit, two
different replicas may have a mutual overlap $q_0= 0$ with probability
m, and $q_1 $ with probability $1-m$.  The Edwards-Anderson order parameter
$q_{EA}(T)$ is equal to $q_1$.  
Note that the shape of $P(q)$ is only sensitive to
the thermodynamics, and for $T_s<T<T_d$, $P(q)=\delta(q) $ like in the
the RS phase. 

The ROM is strongly discontinuous: the value of $q_1$ jumps sharply
from 0 to a value close to 1 at the static transition (see
figure~\ref{fig:011}) below. This, together with the wide separation
between $T_s$ and $T_d$, are strong cases to use the ROM, as compared to
the p-spin model~\cite{BIL} or the Potts
glass~\cite{BRA1,BRA2,BRA3}. Both these models have unfortunately very
close static and dynamic transition temperatures.

\subsection{Numerical method}

Let us start by giving some details about our simulations. We study
systems with $N = 32, 64, 128$ and $256$ spins.  We thermalize
the system using the parallel tempering optimized Monte Carlo
procedure~\cite{Marinari:1992qy,Tesi:1996ud,Hukusima:1996yq,Lyubartsev:1992uq},
with a set of 100 temperatures in the range $[0.08,0.352]$ ($\Delta T =
0.002$ for $T<0.2$ and $\Delta T=0.004$ for $T>0.2$), except for the
largest system where a smaller set of 35 temperatures has been used,
in order to save computer time (with $0.12\leq T \leq 0.28$ and
$\Delta T=0.008$).  These parameters have been chosen empirically and
no claim is made that they are optimal.  As usual, the program
simulates the independent evolution of two clones/replicas, in order
to compute the static overlap $q$. We perform $15 \ 10^6$ parallel
tempering iterations (one iteration consists of one Metropolis sweep
of all spins, followed by one tempering update cycle of all pairs of
successive temperatures). The second half of the equilibration
procedure is used to measure the static quantities. The results of
this procedure are presented in the following subsections (see
subsections~\ref{subsec:3-1} and~\ref{subsec:3-2}).

We then store the final hopefully well-equilibrated
configurations. Due to the difficulty to equilibrate large systems, we
restrict ourselves to $T>T_d$ to study the finite size dependence of
the dynamics.  We also restrict the number of temperatures to 25,
equally distributed between 0.178 and 0.29 ($\Delta T=0.008$), for all
system sizes.  We then perform $5 \ 10^5$ pure Metropolis sweeps and
measure $q_d(t)$, the overlap between the well equilibrated initial
configuration, and the configuration at time $t$.  We then perform $5
\ 10^5$ parallel tempering iterations in order to have a new, well
de-correlated, starting point, and repeat the procedure 100
times. This gives us for each disorder samples $N_{ther}=200$
thermally independent estimates of the dynamical overlap $q_d(t)$,
this is large enough to obtain a reliable estimate of the non-linear
susceptibility $\chi_4$.  The number of disorder samples $\omega$  is
$N_{dis}=500$ for all sizes.

Let us mention that we have also studied the model obtained by
projecting the elements of the matrix $J$ to $\pm 1$. The motivation
was that in this case one could use an efficient multi-spin coding
technique~\cite{Jacobs:1981uq}. Unfortunately, the resulting model
displays a very different physics, with $\infty$-RSB breaking, and is
accordingly not suitable for our purpose. This is an illustration of
the fragility of the ROM model with respect to perturbations.

\subsection{\label{subsec:3-1}Relaxation and equilibration tests}

Disordered systems are notoriously difficult to simulate, and it is
crucial to ensure good quality sampling, and in particular good
thermalization. There is unfortunately no full-proof heuristics for
this purpose. 

\begin{figure}[phtb]
\begin{center}
\includegraphics[width=300pt,angle=0]{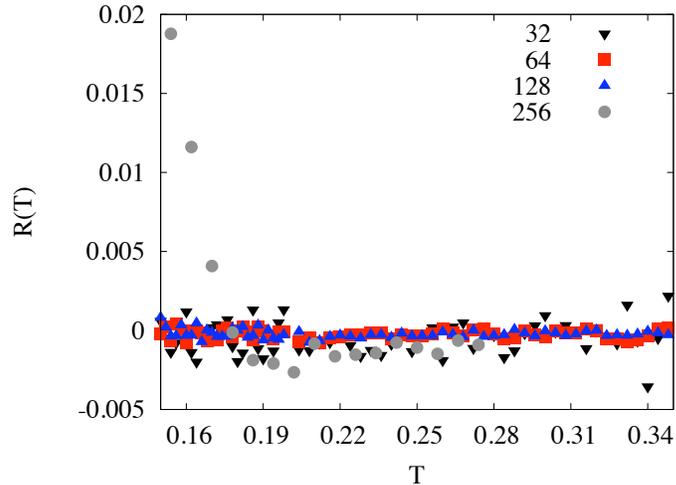}
\caption{Fluctuation-dissipation ratio $R(T)$ (see equation~\ref{R} for the
precise definition) as a function of the temperature, for $N=32$ up to
$N=256$. We note that for all vales of $N$, $R(T)$ fluctuates around
zero, namely  the systems are well equilibrated, except for the largest size ($N=256$) where the system fall
out of equilibrium below $T_d\simeq0.177$. }\label{fig:002}
\end{center}
\end{figure}

The heuristics we use for this simulation is to check that the
fluctuation-dissipation relation, relating the specific heat to the
variance of the internal energy, is satisfied.  This is a very
stringent test (see e.g.~\cite{Krauth:uq}) and we have checked in the
case of the SK model that it is consistent with other methods used in
the literature.  We define the ratio $R(T)$ as:
\begin{eqnarray}
R(T)=\Biggl({\frac{d\overline{\langle e
      \rangle}}{dT}-N\frac{\overline{\langle e^{2} \rangle - \langle e
      \rangle^{2}}}{T^{2}}}\Biggr)\Biggl({\frac{d\overline{\langle e
      \rangle}}{dT}}\Biggr)^{-1},
     \label{R}
\end{eqnarray}
where $e$ is the energy density. $R(T)$ vanishes when the
configurations are well sampled.  Our data for $R(T)$ (with $e$ as
measured in the second half of our thermalization runs) as a function
of $T$ can be found in figure~\ref{fig:002} for system sizes $N=32$ up
to $256$.
The results are very satisfactory for all temperatures up to $N=128$.
For the largest system size however, the sampling is clearly not good
enough below $T_d$, in spite of intensive numerical efforts. This is
to be contrasted to the cases of the SK and Potts glass models where
values of $N$ up to a few thousands can be handled with the parallel
tempering algorithm~\cite{Billoire:2002fj,BRA1}.

\begin{figure}[phtb]
\begin{center}
\includegraphics[width=220pt,angle=0]{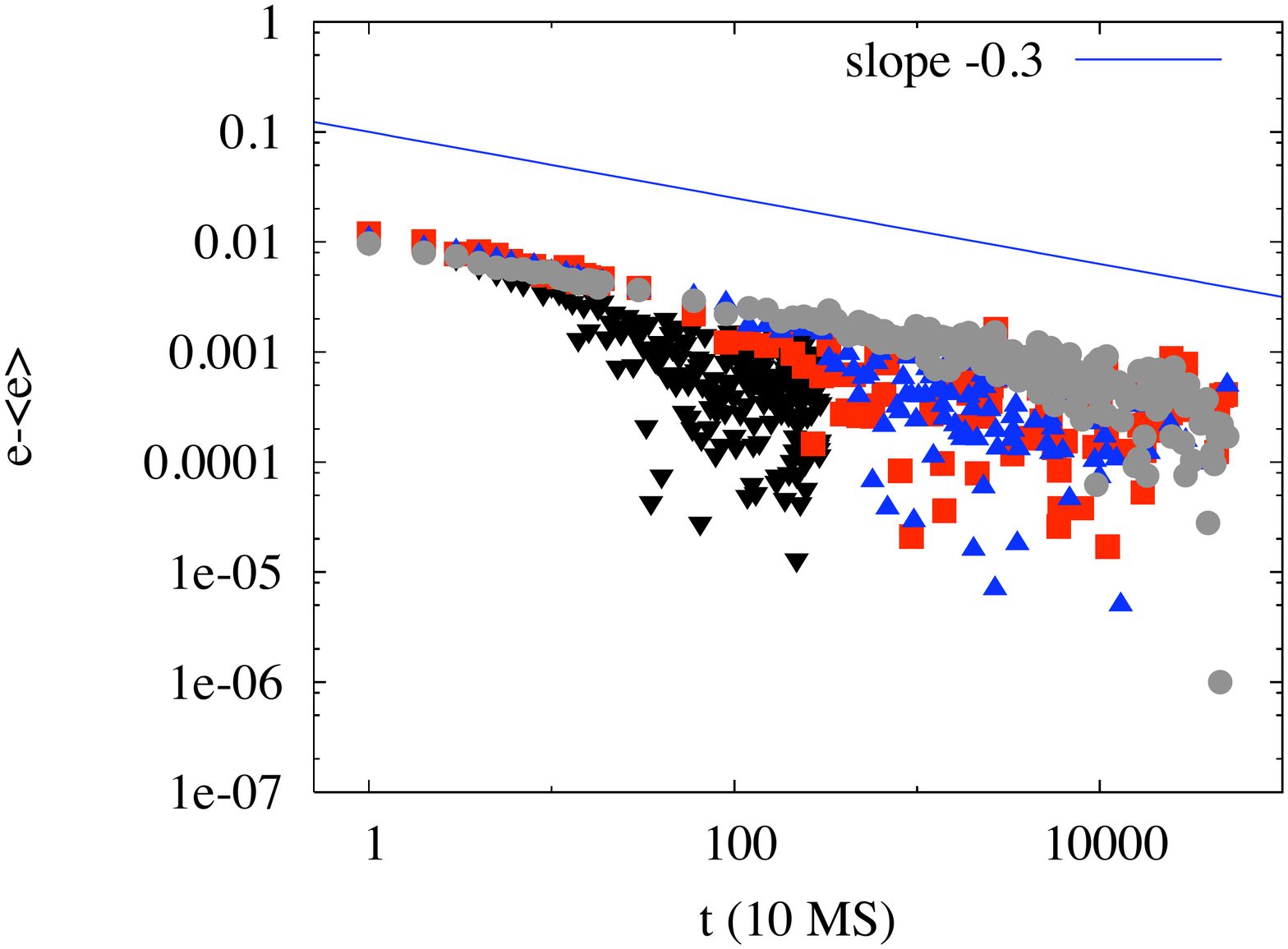}
\includegraphics[width=220pt,angle=0]{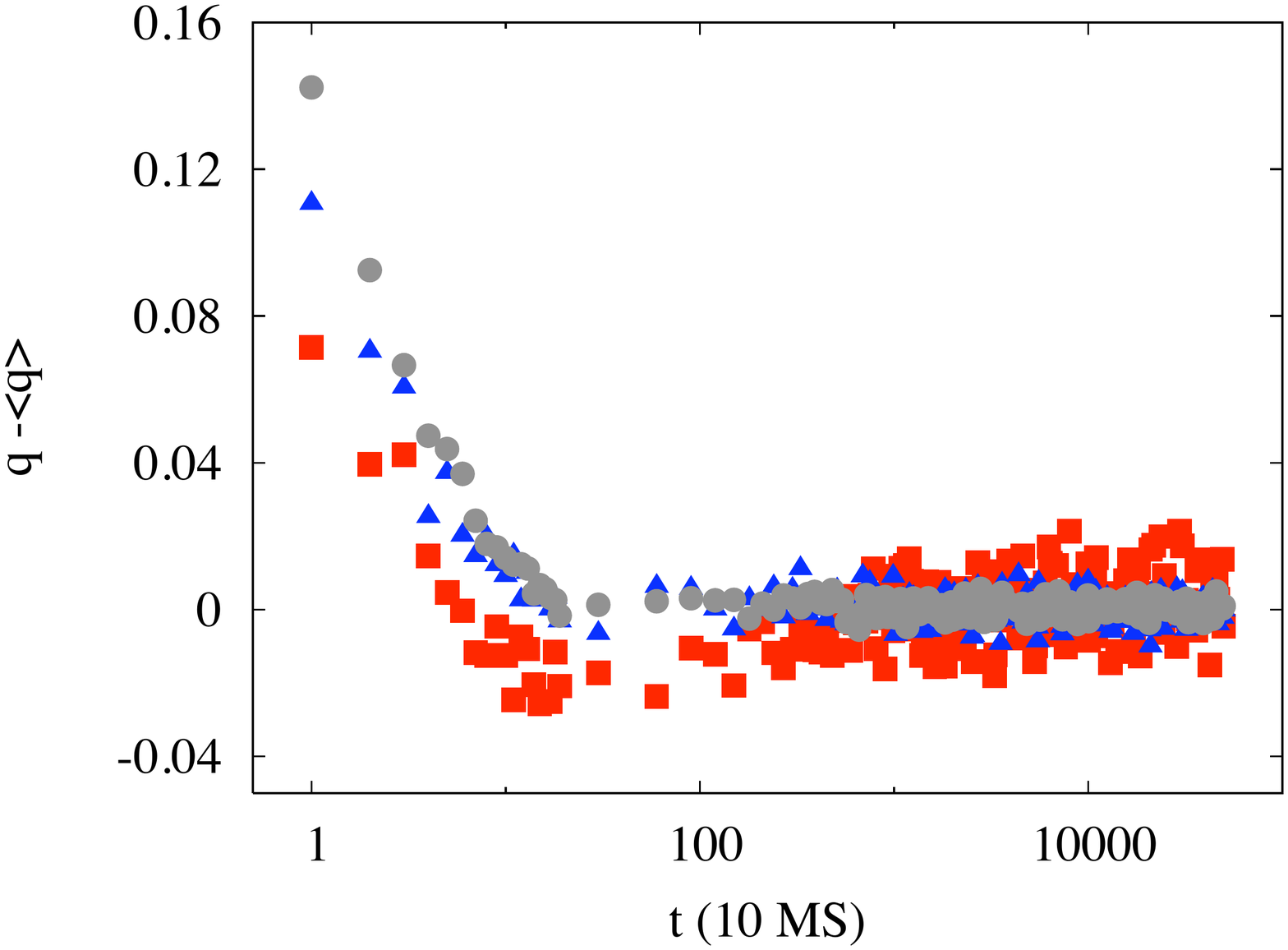}
\caption{Left: Relaxation of the disorder averaged internal energy
  from a random initial configuration, using the parallel tempering
  algorithm, for $N=64,128$ and $256$ at $T=0.178$, just above $T_d$.
  The apparent power law behavior $\overline{e }(t) - e_\infty \propto
  t^{-0.3}$ is only valid in the early time region. One notes that
  even (slightly) above $T_d$, the relaxation is extremely slow and
  equilibration is quite difficult to achieve (one needs a couple of
  $10^5$ Monte Carlo sweeps in order to equilibrate the internal
  energy for $N=256$).  Right: Relaxation of the disorder averaged
  overlap between two clones from disordered initial configurations
  for $N=128$ and $256$ at $T=0.178$. The relaxation is much faster (a
  few hundred Monte Carlo sweeps are enough to equilibrate the
  overlap). We used the same color code of figure~1.}\label{fig:101}
\end{center}
\end{figure}

The modest efficiency of the parallel tempering algorithm when applied to the ROM
can be directly observed by studying how the main thermodynamic
observables reach their equilibrium values.  Starting from a random
initial configuration for the two clones (we take $\sigma_i=1, \forall
i$), we plot the instantaneous value of the considered observable,
averaged over the disorder, in order to tame the fluctuations.  The
results for the internal energy ($\overline{e(t)}$) and for the
overlap between the two clones ($\overline{q(t)}$) are given in
figure~\ref{fig:101}, for $T=0.178 \simeq T_d$. This figure suggest
a power-law behaviour  $\overline{e}(t) - e_\infty \propto
t^{-0.3}$ in the large $N$ limit, with a non uniform
convergence. The smaller the value of $N$, the earlier in time do the
data deviates from the power law behavior. For large $t$, the
relaxation becomes strongly $N$ dependent, smaller systems relax much
faster.

A power-law relaxation of the energy at $T_d$ is in fact expected on
theoretical ground. The value of the exponent was recently conjectured
by A. Lef\`evre (private communication) to be equal to $a$, the MCT
exponent defined above. This is compatible with our finding since, as
we shall find later, $a \approx 0.35$ for the ROM.

The slow relaxation of the internal energy, even for $T>T_d$,
illustrates why this is a hopeless task to simulate the ROM below
(and near) $T_d$ for reasonable system sizes. 
The much faster relaxation of the overlap can be understood 
with the following argument: 
the two clones starts at the top of a very rugged energy landscape. 
After a few sweeps, they have started falling down in directions, or towards, traps whose probable overlap is zero. 
This leads to a fast decorrelation of $\overline{q(t)}$.  Instead,
in order to equilibrate the internal energy, one has to visit many
different traps in order to have a good statistical sampling of
all energy states, which is a much more slow process. 

\begin{figure}[phtb]
\begin{center}
\includegraphics[width=300pt,angle=0]{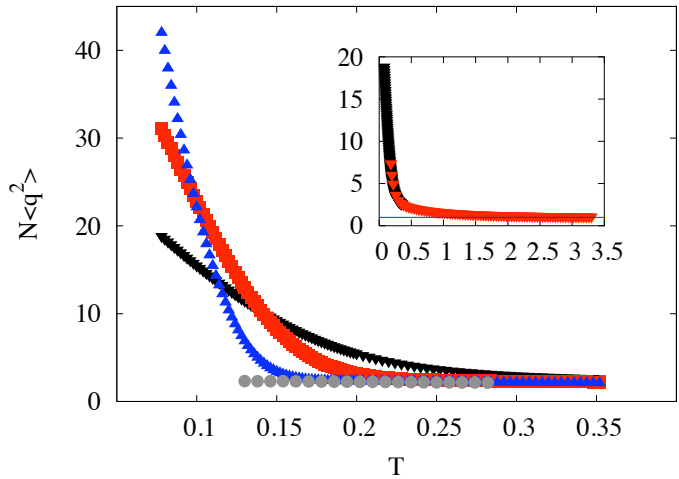}
\caption{Spin-glass susceptibility for the ROM as a function of the
  temperature, for $N=32$ to $256$. One observes (a) strong
  size-dependence effect below $T_d \simeq 0.177$ and (b) an extremely
  slow convergence towards its asymptotic $T \to \infty$ value. In the
  inset $\chi_{SG}^{(N=32)}$ is plotted for high temperatures up to
  $35 T_s$ together with the asymptote $\chi_{SG}^{(\forall N)}=1$.
  This illustrates how ``cold'' the ROM glassy phase is. We used the
  same color code of figure~1.}\label{fig:004}
\end{center}
\end{figure}

Many reasons can be invoked to explain the poor performances of the
parallel tempering algorithm applied to the ROM. The first one
is that the temperature in the interesting region is in fact extremely
low, leading to extremely small Metropolis and exchange acceptance
rates. A way to gauge the smallness of the ROM transition temperature
is to study the spin-glass susceptibility $\chi_{SG}=N\overline{\langle q^2 \rangle}$  as a function of the
temperature (see figure~\ref{fig:004}). When $T\to\infty$,
$\chi_{SG}$ must tend to unity, but strong corrections are 
still present up to $T \approx 10 T_d$. This is expected, since the
leading non trivial $1/T$ correction to $\chi_{SG}$ is the same in the
ROM and SK models, because the variance of the $J_{ij}$ is normalized
to the same value in both cases. However, the critical temperature of
the SK model is $T_c^{SK}=1$, ten times larger than the critical
temperature of the ROM.

A deeper explanation relies on the physics of 1-RSB models, recalled
in the above section. The existence of numerous metastable states in
the range [$T_s$,$T_d$] slows down dramatically the dynamics. Indeed,
in order to have a decent sampling, one should explore a
representative subset of all the metastable states. The complexity of
the ROM has been computed in the thermodynamic limit
in~\cite{PAR3}. It shows a sharp jump from $0$ to a finite value below
$T_d$, and thus the log of the number of states that must be explored
jumps from $0$ to a number of order $N$ at $T_d$.  It does not come as
a surprise that the algorithm fails below $T_d$ even for moderate
values of $N$.  A precise understanding of why the ROM case is so
much difficult than the Potts and p-spin cases is still
lacking. A reasonable conjecture is that this is due to fact that
the overlap value is so close to unity. Another system where this happens, and
where the dynamics is indeed painfully slow, is the Bernasconi model (see \cite{Bernasconi, BM94}).

We note, en passant, that applied to the model with binarized exchange
couplings briefly mentioned above, the parallel tempering works
brilliantly.


\subsection{\label{subsec:3-2} Thermodynamics}

\begin{figure}[phtb]
\begin{center}
\includegraphics[width=300pt,angle=0]{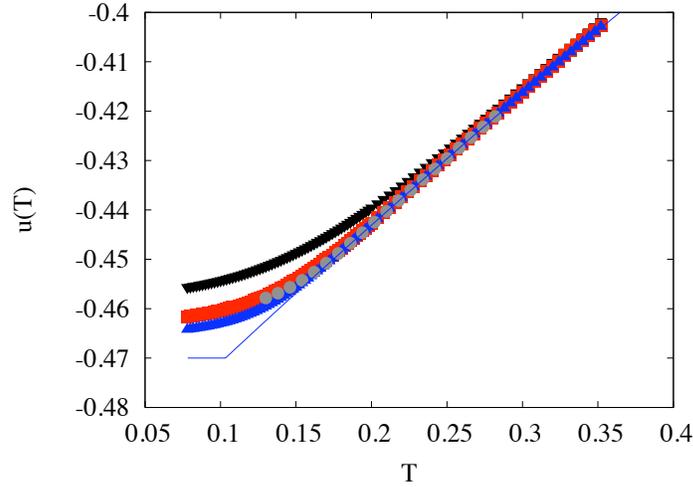}
\caption{Internal energy density $u(T)$ as a function of T for $N=32$
to $N=128$ (top to bottom) and $256$ (points that do not extend below
$T=0.12$), together with the analytical prediction in the infinite
volume limit.}\label{fig:001}
\end{center}
\end{figure}

We show in figure~\ref{fig:001} our data for the internal energy per
spin, defined as $u(T)=\overline{\langle H \rangle}/N$ as a function
of the temperature, together with the theoretical result obtained
in~\cite{LEF1}. For our special choice of $p$, one finds:
\begin{eqnarray}
u(T)&=&\frac{1}{4}\left(T-\sqrt{T^2-\frac{3}{4}T+4}\right) \ , \ T \geq T_s
\ ,\\ u(T)&=&-0.47 \ , \ T<T_s\ .
\end{eqnarray}
The numerical data for $u(T)$ is qualitatively consistent with the
infinite-volume analytical results, up to the finite size corrections,
with the marked exception of the $N=256$ data below $T_d$, which are
at odds with the rest of the picture. This is in agreement with our
previous observation that the $N=256$ systems are not at equilibrium
at low temperature and should be discarded there. The temperature
below which the fluctuation-dissipation relation is violated indeed
roughly coincides with the one below which the numerical values of
$u(T)$ become manifestly wrong.

After discarding the bad data, $u(T)$ converges towards the predicted
 infinite-volume
value, although with marked size-effects below the dynamical
temperature.  The finite size effects at $T=0.08$ are roughly
compatible with a $1/N$ behavior.  A similar observation was made for
the p-spin model in~\cite{Billoire:2002fj}.

\begin{figure}[phtb]
\begin{center}
\includegraphics[width=220pt,angle=0]{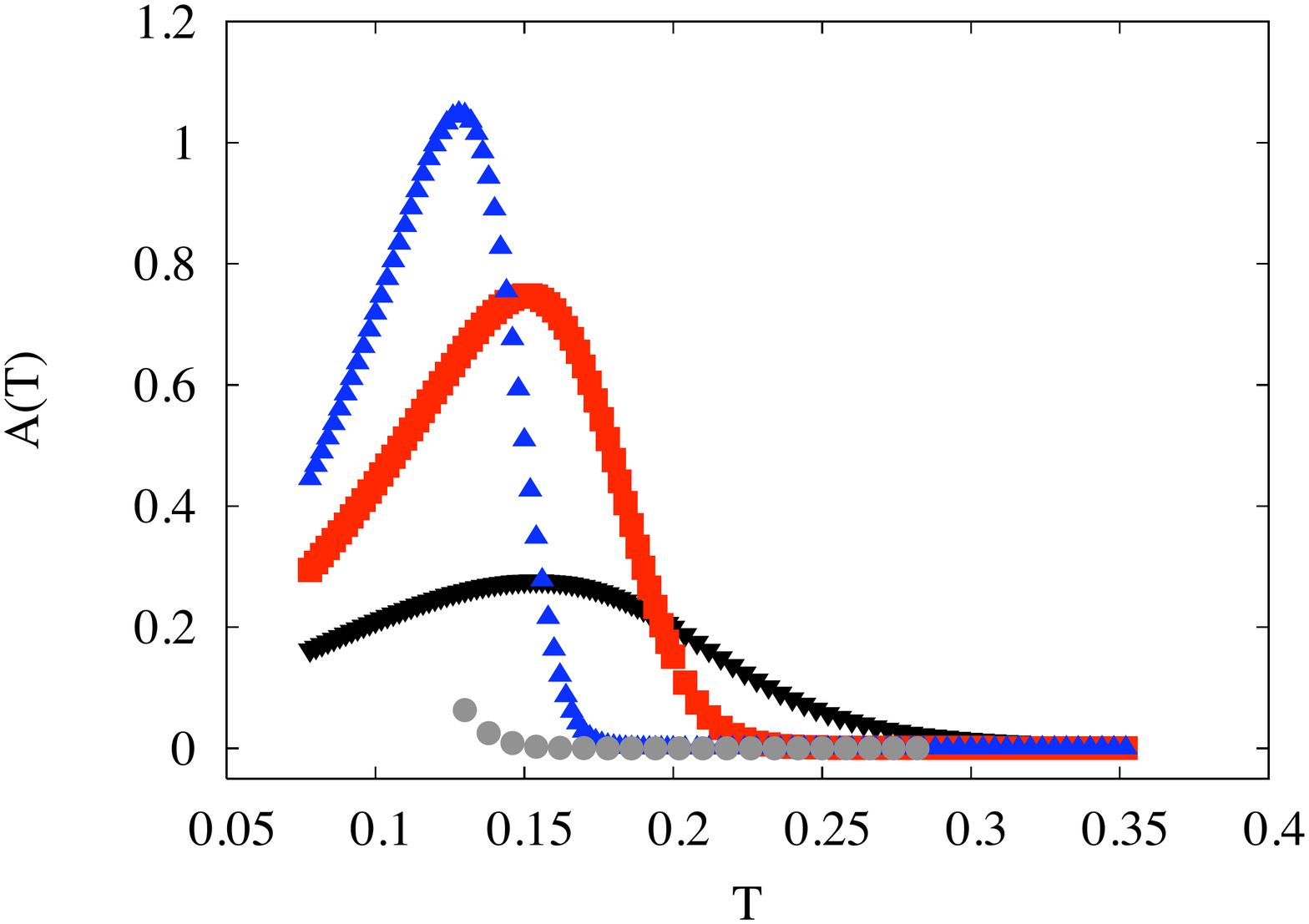}
\includegraphics[width=220pt,angle=0]{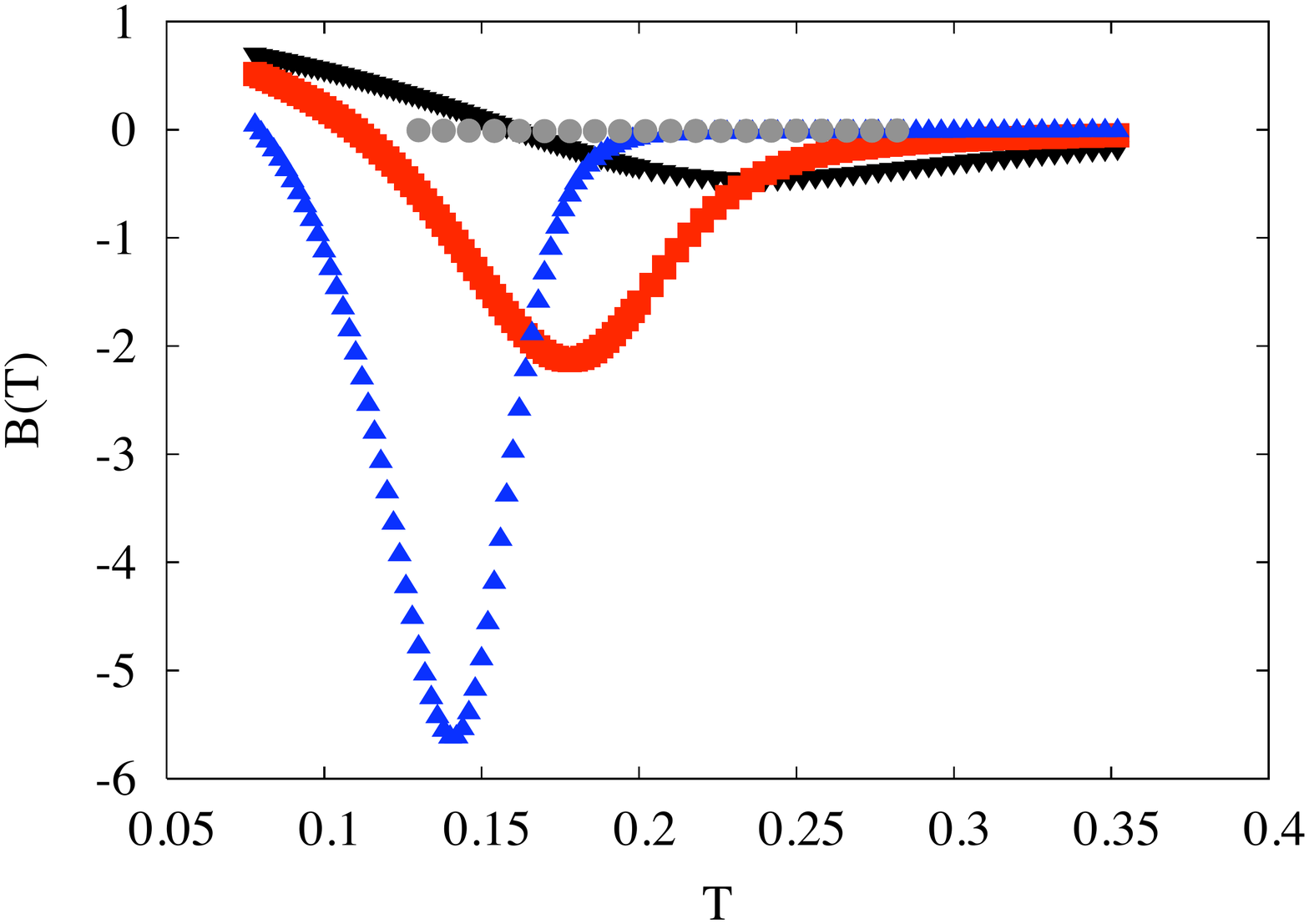}
\caption{The parameter $A$ (left) and the  Binder parameter $B(T)$  (right)
as a function of $T$, for$N=32$ to $256$. We used the same color code of the previous figures.}\label{fig:006}\label{fig:007}
 \end{center}
\end{figure}

Following~\cite{Picco:2000kx}, we plot in figure
(\ref{fig:007}) the coefficient $A(T)$ defined as:
\begin{eqnarray}
A(T)=\frac{\overline{\langle q^2 \rangle^2}\ -
\overline{\langle q^2 \rangle}^2}{\overline{\langle q^2 \rangle}^2}\ ,
\end{eqnarray}
that signals~\cite{Marinari:1999uq} the onset of the non
self-averaging behavior of $q^2$.  In figure~\ref{fig:006}, we plot
the usual Binder parameter,
\begin{eqnarray}
B(T)=\frac{1}{2}\left(3-\frac{\overline{\langle q^4
\rangle}}{\overline{\langle q^2 \rangle}^2}\right) \ .
\end{eqnarray}

For generic 1-RSB transitions, Picco et al.~\cite{Picco:2000kx} have
argued that these two coefficients are zero for $T > T_s$ (in the
thermodynamic limit), non zero for $T < T_s$ and in fact diverge at
the static transition $T \to T_s^-$.  The fact that the Binder
coefficient is negative when $T \to T_s^-$ is simply related to the
appearance of a peak in $P(q)$ for $q \neq 0$. In the infinite volume
limit, one has:
\begin{eqnarray}
P(q)=(1-m)\delta(q)+m\delta(q-q_1),
\end{eqnarray}
with $m < 1$, and $m \to 0$ when $T \to T_s^-$. Accordingly one has
$B(T)=1/2( 1-1/m) < 0$ in the 1-RSB region.

Our data are in qualitative agreement with the above limiting behavior
(see~\cite{Picco:2000kx,BIL} for a similar numerical analysis in the
case of the p-spin model). Note that for a 1-RSB transition, the
curves of $B(T)$ as a function of $T$ for various values of $N$ do
{\it not} cross at a universal point, at variance with usual phase
transitions (including $\infty$-RSB transitions).

\begin{figure}[phtb]
\begin{center}
\includegraphics[width=220pt,angle=0]{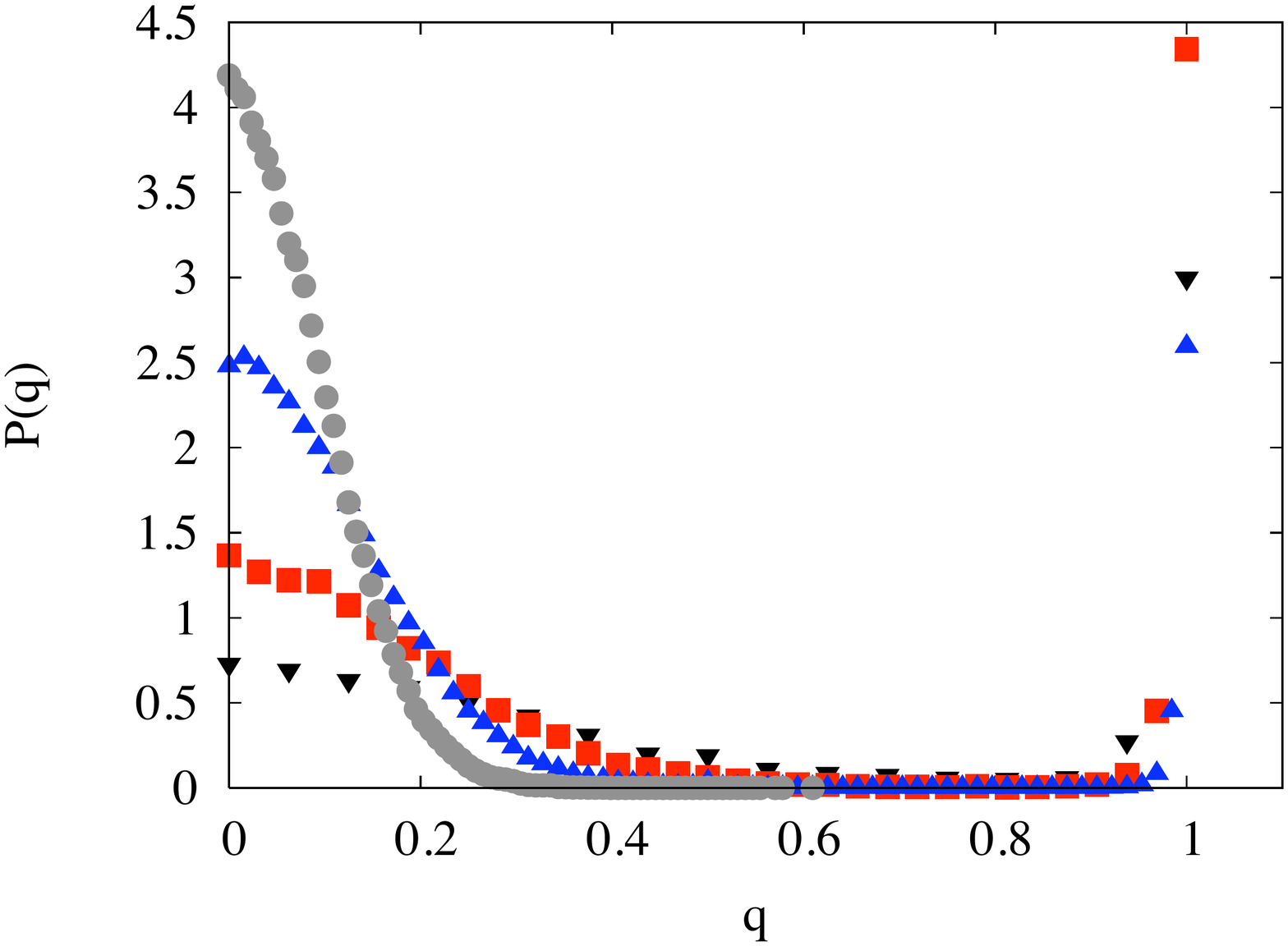}
\includegraphics[width=220pt,angle=0]{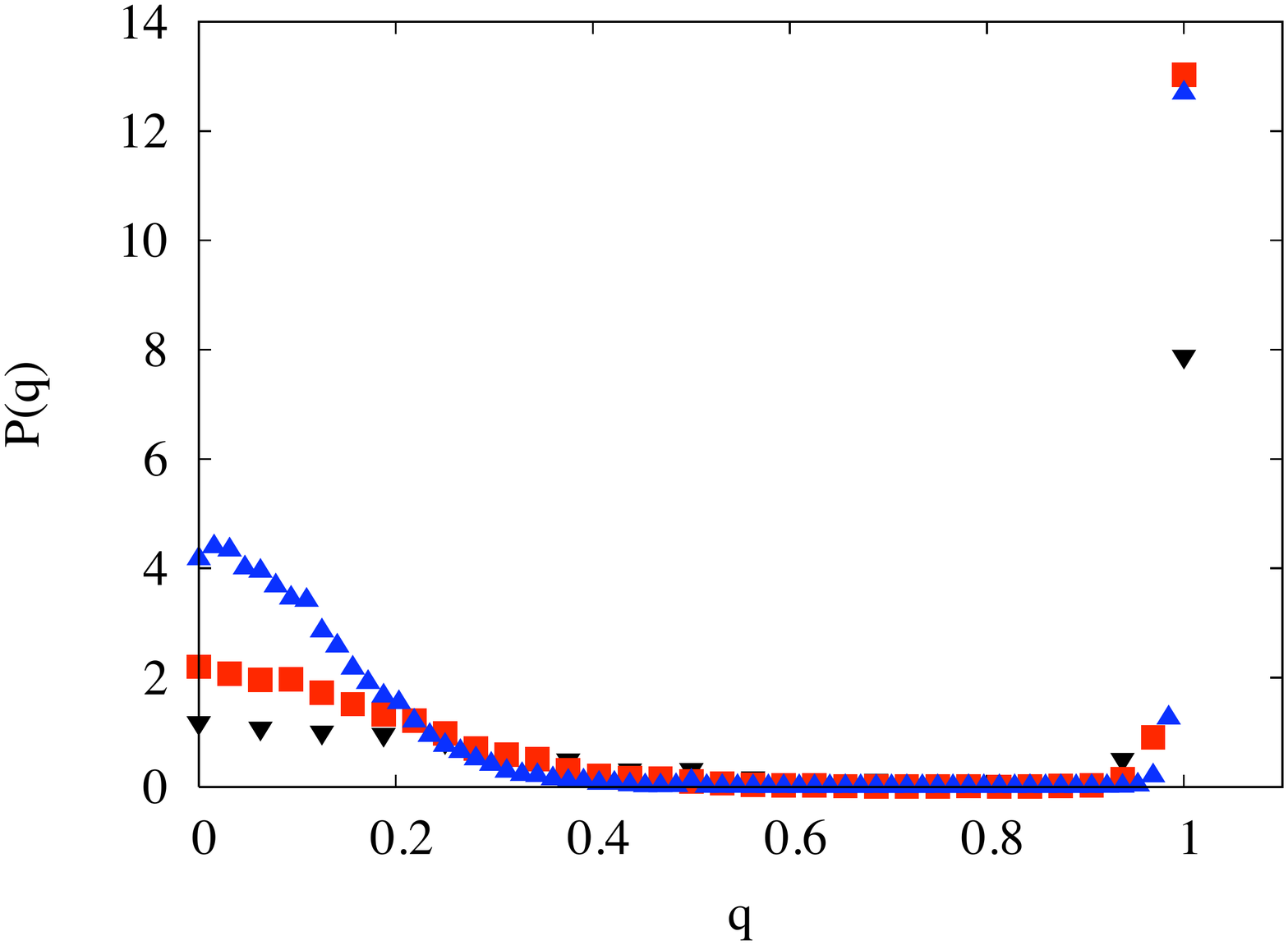}
\caption{Left: $P(q)$ for $T_s < T=0.154 < T_d$.. Right:$P(q)$ for
  $T=0.078 < T_s$. Both for $N=32$ to $128$. We used the same color
  code of the previous figures. }
\label{fig:011}\label{fig:013}
\end{center}
\end{figure}

The overlap probability distribution $P(q)$ is found to have the shape
corresponding to a 1-RSB phase transition, with one peak around $q=0$
and, below $T_s$, another peak around $q=q_1$ ($q_1$ is close to one
in our specific case).  With $N$ finite, both peaks have a non zero
width, as always.  It turns out that the $q=0$ peak is much broader
that the peak at $q=q_1$.  Above $T_s$, our data shows a spurious peak
centered at $q \simeq 1$.  This peak however corresponds to an
unstable thermodynamic phase and decays sharply with the size of the
system (see figure~\ref{fig:011}) as it should.  As shown in figure
(\ref{fig:013}), the peak centered at $q\simeq1$ becomes extremely
sharp for $T < T_s$ .

\subsection{\label{subsec:3-3} Thermodynamics: Conclusion}

We have thus been able to confirm numerically the main replica
predictions for the ROM: the energy as a function of the temperature
freezes at the static transition, and the order parameter is strongly
discontinuous there.  This could be related to the fact that the static
transition temperature is very small, compared, for example, to the SK
model. Correspondingly, it is very hard to equilibrate the system
despite intensive numerical efforts. Systems with $N=256$  did
not reach equilibrium below the dynamical transition.

\section{\label{sec:dyn}Numerical Simulations: dynamical behavior}

We now present our numerical study of the {\it equilibrium} dynamics
of the ROM. On theoretical grounds, and in view of the above results
on the statics of the model, the dynamics of the ROM should be
described by the Mode-Coupling Theory, at least in a Landau sense, i.e. the MCT power laws are expected to be valid and the actual values of the exponent, although not universal, are constrained to verify equations~\ref{eq:exp} and\ref{eq:tau}, see\cite{ABB})  
We compare our data with the
predictions of MCT both for the two-point and four-point correlation
functions.  The results are very puzzling at first sight. We will 
show in the following sections that a detailed understanding of
preasymptotic corrections and finite size effects is required in order
to rationalize our numerical results.

\subsection{Dynamic scaling and comparison with MCT}

\begin{figure}[phtb]
\begin{center}
\includegraphics[width=220pt,angle=0]{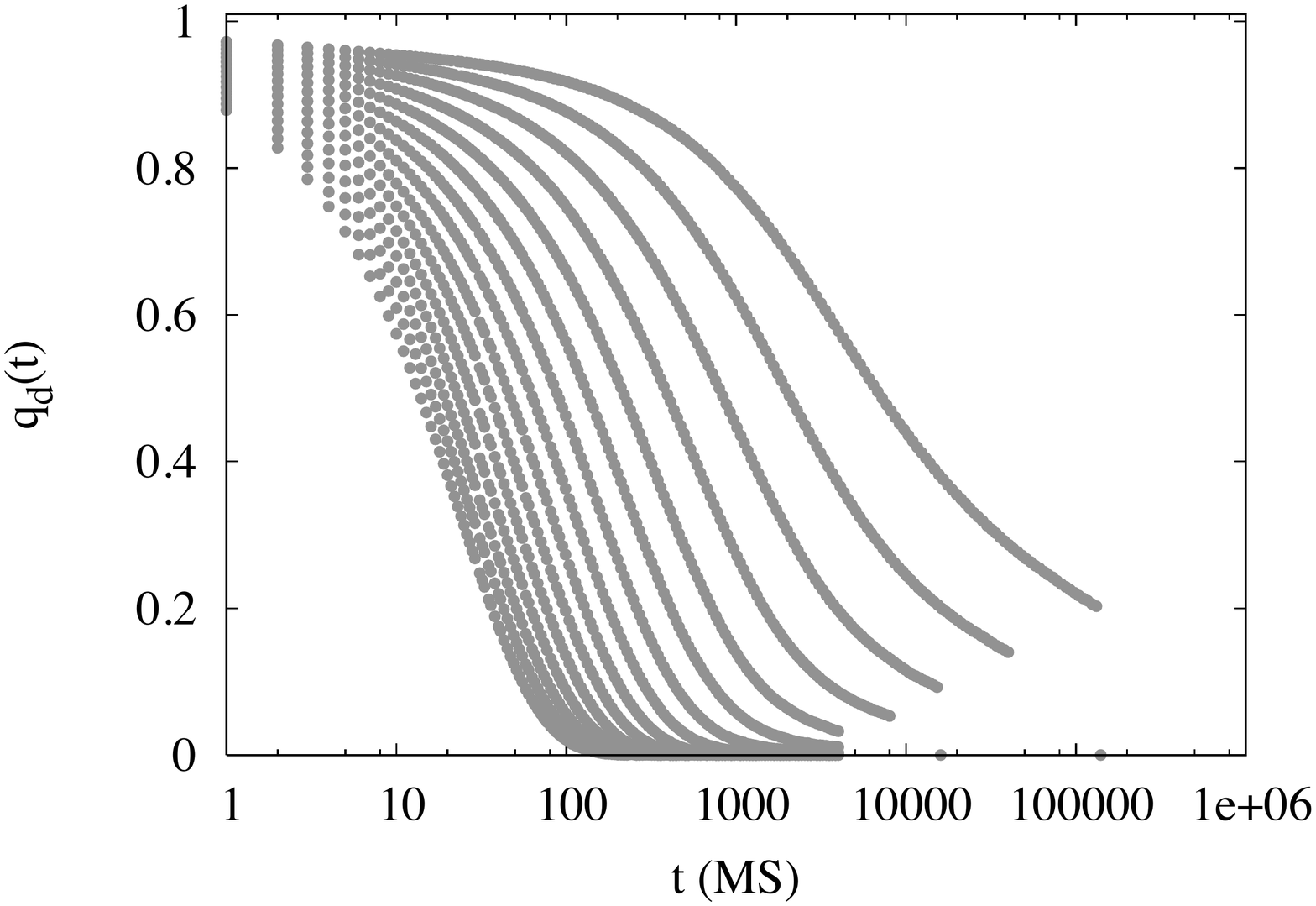}
\includegraphics[width=220pt,angle=0]{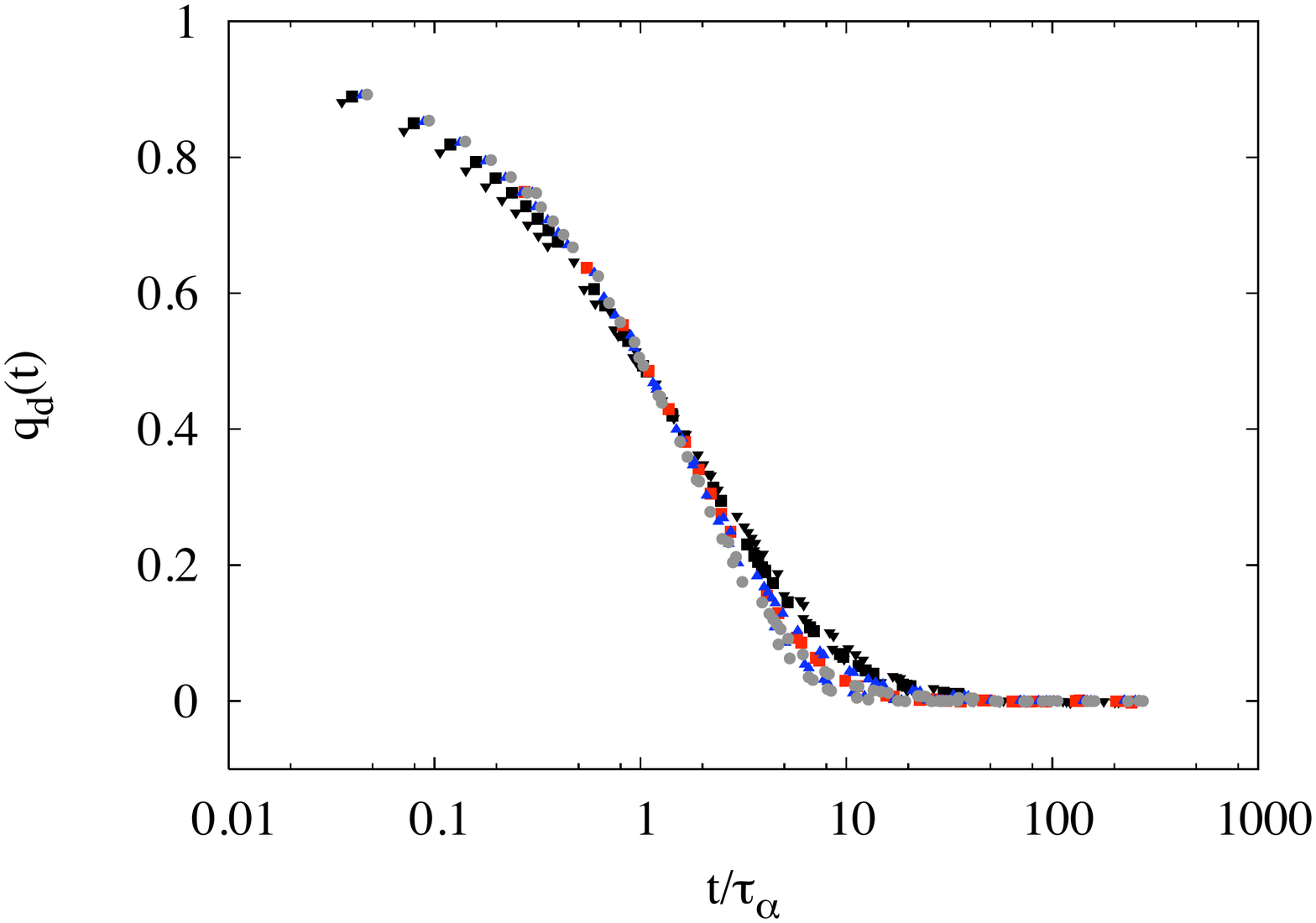}
\caption{Left: Dynamical overlap $q_d(t)$ for the largest system ($N=256$),
and for various temperatures from $T\simeq T_d$ (upper curve) to
$T=1.65\ T_d$ (The temperature step is such that $\Delta T/T_d = 5\%
$).Right: Test of Time-Temperature superposition for $q_d(t)$ for all
sizes ($N=256$ grey circle, $128$ blue upper triangle, $64$ red
squares and $32$ black lower triangle) and various temperatures from
$T=1.25\ T_d$ to $T=1.65\ T_d$. One remarks clearly the decay of
finite-size corrections and the convergence to a limiting
curve.}\label{fig:4A1}\label{fig:4A2}
\end{center}
\end{figure}

We first focus on the dynamical overlap $q_d(t)$ (defined in
equation~\ref{qddt}).  Our data show a plateau in $q_d(t)$, whose extension
increases by lowering the temperature, see Fig~\ref{fig:4A1}.

The value of the dynamical overlap on the plateau is close to the
infinite-volume limit of the Edwards--Anderson parameter
($q_{EA}(T_d)\approx 0.955$), as it should. After a single Monte Carlo
sweep, $q_d(t)$ has already decayed to the plateau value. Therefore we
cannot observe the early $\beta$-regime.  This is an unfortunate
drawback of our choice of the ROM with parameter $p=13/32$.  We
however do see the $\alpha$ regime in full glory.  Our results are
quite different from those obtained for the Potts
model~\cite{BRA1,BRA2,BRA3}, where no plateau was observed in $q_d(t)$
for systems with up to $N=2560$.

In order to be quantitative, we define the time scale $\tau_\alpha(T)$
(see equation~\ref{tts}) as the time needed to reach the value
$q_d(\tau_\alpha)=1/2$. If Time-Temperature superposition (TTS) holds,
the precise definition used is irrelevant. We have
checked that the $q_d(t)$'s plotted as a function of $t/\tau_\alpha(T)$
approximately collapse on a unique scaling curve for the largest
system sizes, but for temperatures not too close to $T_d$, see Fig
\ref{fig:4A2}.

\begin{figure}[phtb]
\begin{center}
\includegraphics[width=220pt,angle=0]{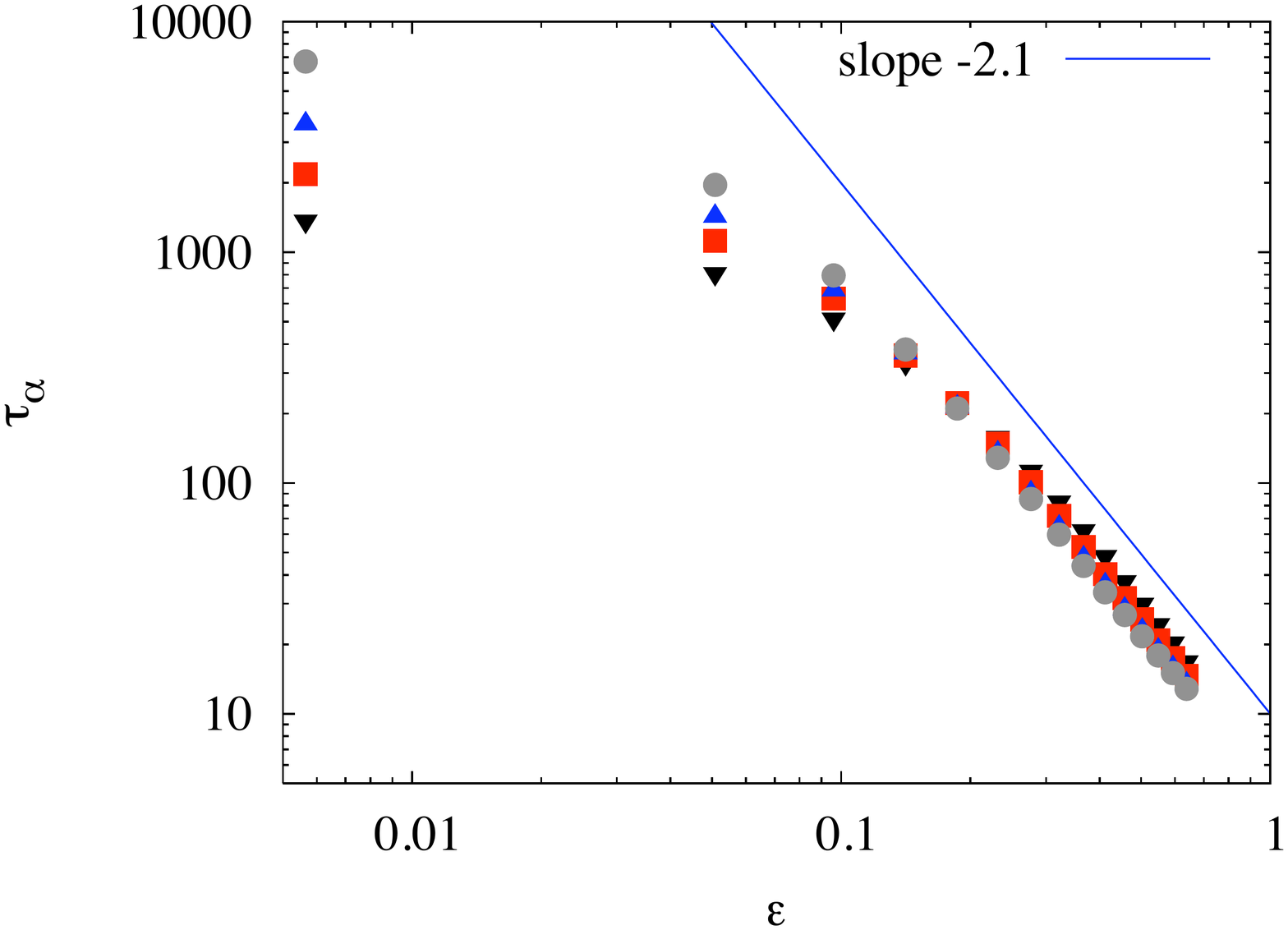}
\includegraphics[width=220pt,angle=0]{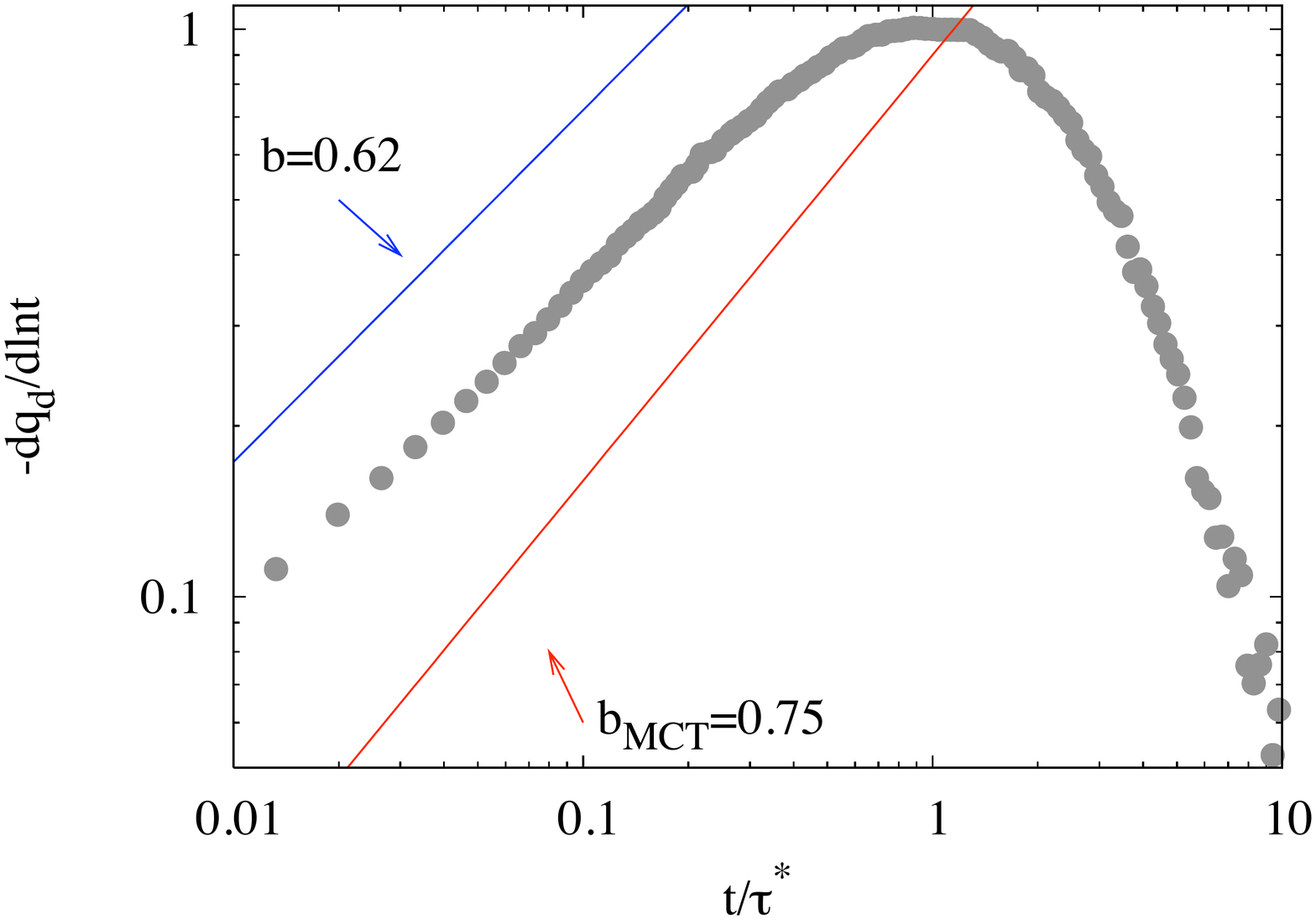}
\caption{Left: Relaxation times $\tau_\alpha$ versus $\epsilon=(T-T_d)/T_d$,  for
different system sizes (same color code of previous figures). For $\epsilon > 0.2$, i.e. $T > 1.2\ T_d$, an
$N$-independent regime is observed, with $\tau_\alpha \propto
\epsilon^{-\gamma}$ and $\gamma \approx 2.1$, over two time decades. 
Right: Determination of the $b$-exponent of the von Schweidler law
through the derivative of the overlap, $-dq_d/d\ln t \propto t^b$.
(Here $N=256$, $T=0.226$). The two straight lines correspond to
$b=0.62$ and $b_{MCT}=0.75$ respectively. }\label{fig:4A3}\label{fig:4A4}
\end{center}
\end{figure}

However, as $T$ gets closer to $T_d$, TTS appears to breakdown:
instead of approaching a universal scaling curve, the $q_d(t)$'s are
more and more stretched as the system size increases. This is a priori
surprising since TTS should work better close to the MCT
transition. But we find that close to $T_d$, finite size effects
become important. This is also revealed by the behavior of
$\tau_\alpha(T)$ as a function of $\epsilon=(T-T_d)/T_d$, see Fig
\ref{fig:4A3}. A MCT power law fit, $\tau_\alpha\propto \epsilon
^{-\gamma}$, accounts reasonably well for the regime $\epsilon > 0.2$
where finite size effects are small, and yields $\gamma \approx 2.1$.
MCT also makes detailed prediction about the form of the scaled
relaxation function, as we discussed in the introductory
sections. Unfortunately the power-law predicted in the early $\beta$
regime is inaccessible because the plateau is too close to
unity. However, we can check the (von Schweidler) power-law associated
to the late $\beta$ regime. In order to do so in a way that does not
require a precise determination of the plateau value $q_{EA}$, we plot
in figure~\ref{fig:4A4} $-dq_d/d\ln t$ for our largest system size
$N=256$ at the lowest temperature at which we do not have substantial
finite size effects. According to MCT, this quantity should increase
as $t^b$ when $\tau_\beta \ll t \ll \tau_\alpha$. From the power law
fit shown in figure~\ref{fig:4A4} we determine $b\approx 0.62 $. But
within MCT the values of $b$ and $\gamma$ are not independent (see
equations~(\ref{eq:exp},\ref{eq:gamma}) above). The value of $b$
corresponding to $\gamma \approx 2.1$ is found to be $b_{MCT}\approx
0.75$ which is distinctly too large compared to our data (see Fig
\ref{fig:4A4}). This is a second puzzling result since MCT predictions
are expected to apply to the ROM dynamics close to $T_d$.

\begin{figure}[phtb]
\begin{center}
\includegraphics[width=220pt,angle=0]{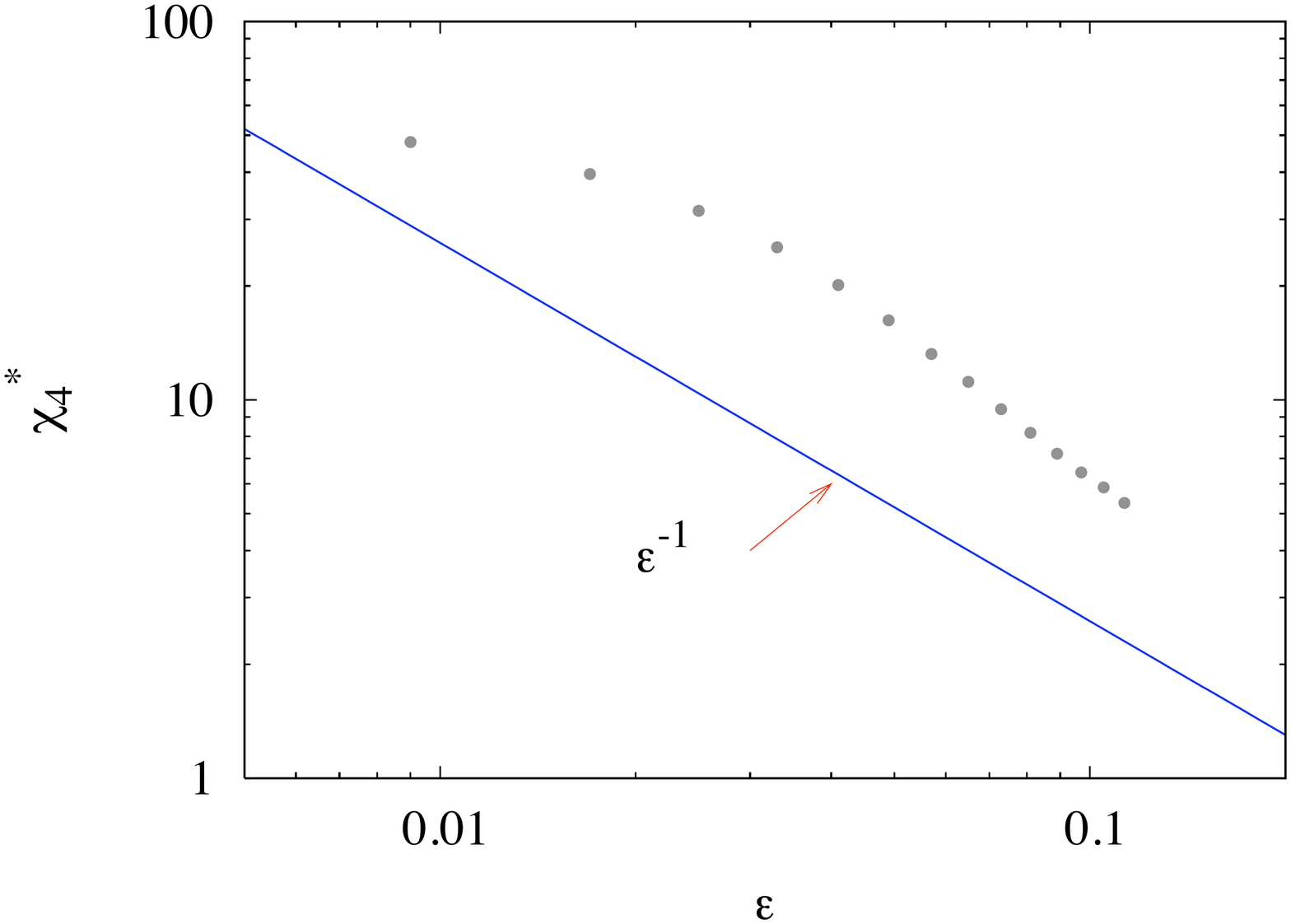}
\includegraphics[width=220pt,angle=0]{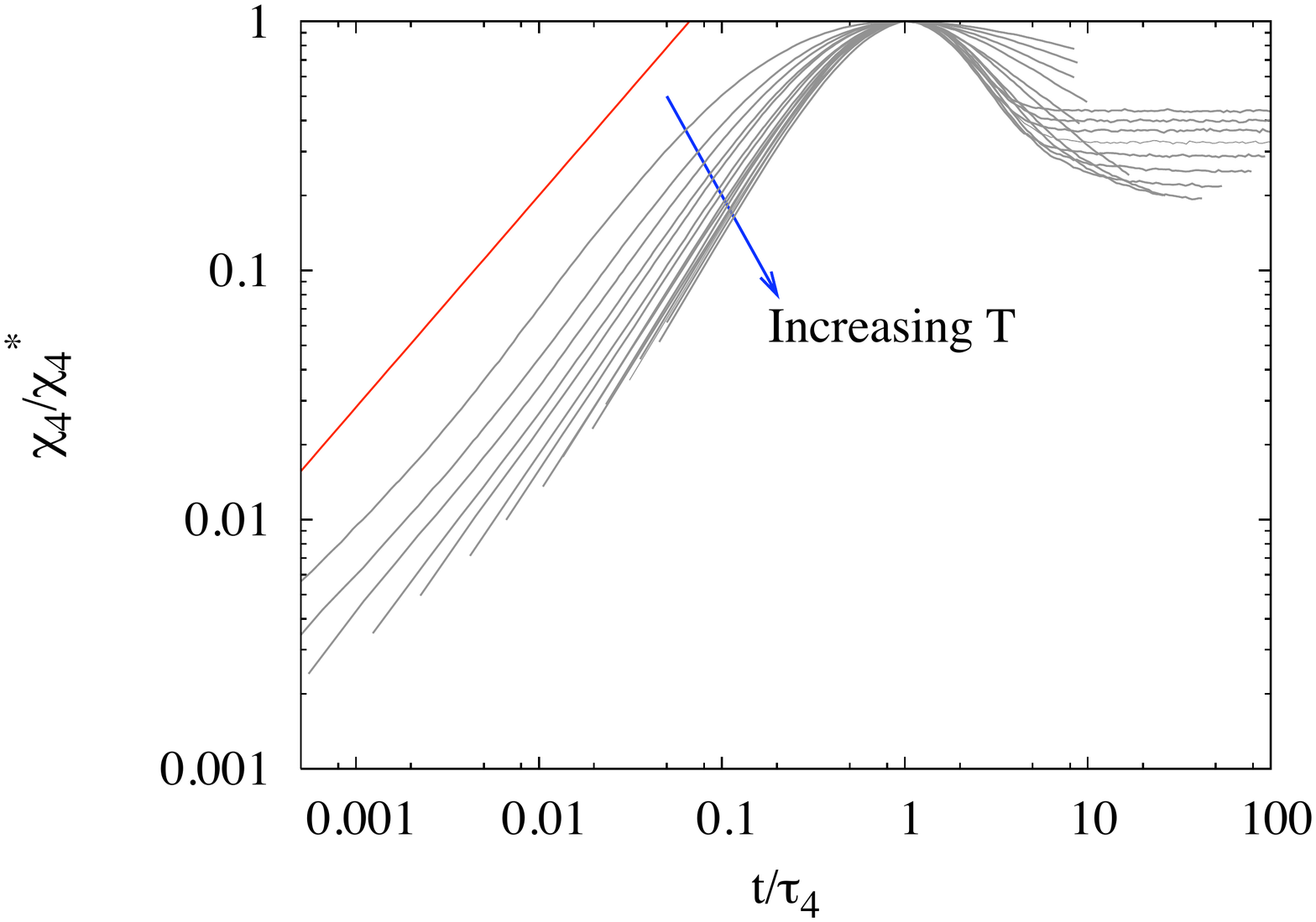}
\caption{Left: Maximum of $\chi_4$ versus $\epsilon$. The MCT
prediction would be $\chi_4^*\propto 1/\epsilon$. Right: Scaling plot for $\chi_4$ (here
N=256). The straight line corresponds to a power law with an exponent $0.9$. }\label{fig:4A5}\label{fig:4A6}
\end{center}
\end{figure}

\begin{figure}[phtb]
\begin{center}
\includegraphics[width=300pt,angle=0]{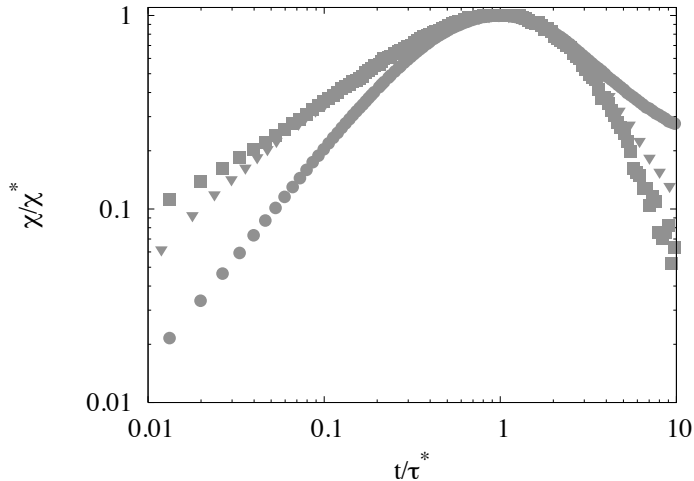}
\caption{ Comparison of three susceptibilities: $\chi_4$ (circles),
$-dq_d/d\ln t$ (squares) and $-dq_d/dT$ (triangles). Here, $N=256$ and
$T=0.226$.  Each susceptibility has been rescaled with its
own $\chi^*$ and $\tau^*$. All three quantities should grow with the same initial
power-law behavior within MCT. This is clearly not the case for
$\chi_4$(t).}\label{fig:4A7}
\end{center}
\end{figure}

Other puzzling features emerge from the analysis of the dynamic
susceptibility $\chi_4(t)$, as defined by equation (\ref{eq:defchi4}).  A
plot of the peak value $\chi_4^*$ as a function of $\epsilon$ for our
largest system size shows (in the regime without finite size effects)
a power law behavior compatible with the MCT prediction
$\chi_4^*\propto 1/\epsilon$ (see figure~\ref{fig:4A5}). However,
figure~\ref{fig:4A6} shows that there is no collapse of
$\chi_4(t)/\chi_4^*$ plotted as a function of $t/\tau_4$, where
$\tau_4$ is such that $\chi_4^*\equiv\chi_4(t=\tau_4$)\footnote{We
checked that $\tau_4$ and $\tau_\alpha$ are approximatively
proportional.}, at variance with the MCT prediction for $t <
\tau_4$~\footnote{For large times scaling is not expected, 
since in this limit $\chi_4(t) \to \chi_{SG}$ and this
has a finite nonzero limit as $N\to\infty$ and $T\to T_d$, 
contrary to $\chi^*_4$ that is expected to diverge.}.
Before the peak, $\chi_4(t)$ appears to grow as $t^{b_{4}}$ with
$b_{4} \approx 0.85$. Such a power-law increase is again predicted by
MCT, but one should find $\chi_4(t) \propto -dq_d/d\ln
t$~\cite{BB-EPL,IMCT}, i.e. $b_4=b \approx 0.62$.  In fact, MCT also
predicts that $\chi_4(t) \propto -dq_d/dT$~\cite{BBBKKR}. We show in
figure~\ref{fig:4A7} these three different quantities in log-log
plot. Although $-dq_d(t)/dT$ and $-dq_d/d\ln t$ are indeed similar,
$\chi_4(t)$ does not conform to expectations. 

In conclusion: although some of the MCT predictions are quantitatively
obeyed, other important ones are clearly violated. The solution of
this conundrum is that the values of $\epsilon$ used above are
actually not small enough to be in the asymptotic regime where MCT
predictions hold. As we shall show in the next section, these
predictions are only valid in a {\it surprisingly small region} close
to the transition. So why not work closer to $T_d$? The next problem
we will have to deal with (section~\ref{sec:4}) is finite size effects,
that become large close to $T_d$. Therefore, only after a very careful
finite size analysis can one conclude on the compatibility between the
MCT predictions and the numerical behavior of a model that is in
principle exactly described by MCT! We will show how difficult this
program turns out to be for the ROM. This sheds considerable doubt on
the precise, quantitative comparison between experimental data and
MCT, since these problems should show up in these cases as well.

\subsection{MCT critical properties and preasymptotic corrections}

At this stage, it is important to have a reliable reference point to
which we can compare our numerical results. For this we choose to
study in detail the Leutheusser integro-differential equation for the
correlation function~\cite{Leutheusser:1984uq}, which comes out of the
schematic version of MCT with a so-called quadratic kernel:
\begin{eqnarray}
\ddot \Phi(t)+\Omega \dot \Phi(t)+\Phi(t)+4\lambda \int_0^t d\tau
\Phi^2(\tau) \dot \Phi(t-\tau)=0,\hskip 0.5cm \Phi(0)=1
\end{eqnarray}
In the above equation, $\Phi(t)$ is the correlation function and plays
the role of $q_d(t)$ above, and $\lambda$ is the coupling constant
that measures the strength of the feedback effects at the heart of the
MCT transition. Remarkably, the equation above is also the one governing
the evolution of the correlation function for the $p=3$ mean field disordered
p-spin model \cite{Cugliandolo:ao}.  It can be analyzed mathematically, and all
the results quoted in section~\ref{sec:2} can be shown to hold exactly
in the limit $\lambda \to \lambda_d =1$.  In particular, the model is
ergodic for $\lambda<1$, where $\lim_{t\to\infty} \Phi(t)=0$, and
develops power-law regimes with exponents $a$ and $b$ given by
(equation~\ref{eq:exp}):
\begin{eqnarray}
\label{gammaeq2}
\frac{\Gamma^2(1+b)}{\Gamma(1+2b)}=\frac{\Gamma^2(1-a)}{\Gamma(1-2a)}=\frac{1}{2},
\end{eqnarray}
leading to $a\approx 0.315$, $b=1$, $\gamma\approx1.765$.

The Leutheusser equation can be solved numerically for arbitrary large
values of $t$, using for example the algorithm of~\cite{Fuchs:1991sf}.
In what follows we fix $\Omega=1$, and neglect the $\ddot \Phi(t)$
term, as usually done. In order to compare directly with the ROM data
above, we have computed $\Phi(t)$ for values of $\lambda$ that are at
the same relative distances from the critical point as our ROM data,
for temperatures $T=0.178, 0.186,\ldots$, and $0.250$. (We recall that
$T_d=0.177$ for the ROM.)

\begin{figure}[htb]
\centering 
\includegraphics[width=220pt,angle=0]{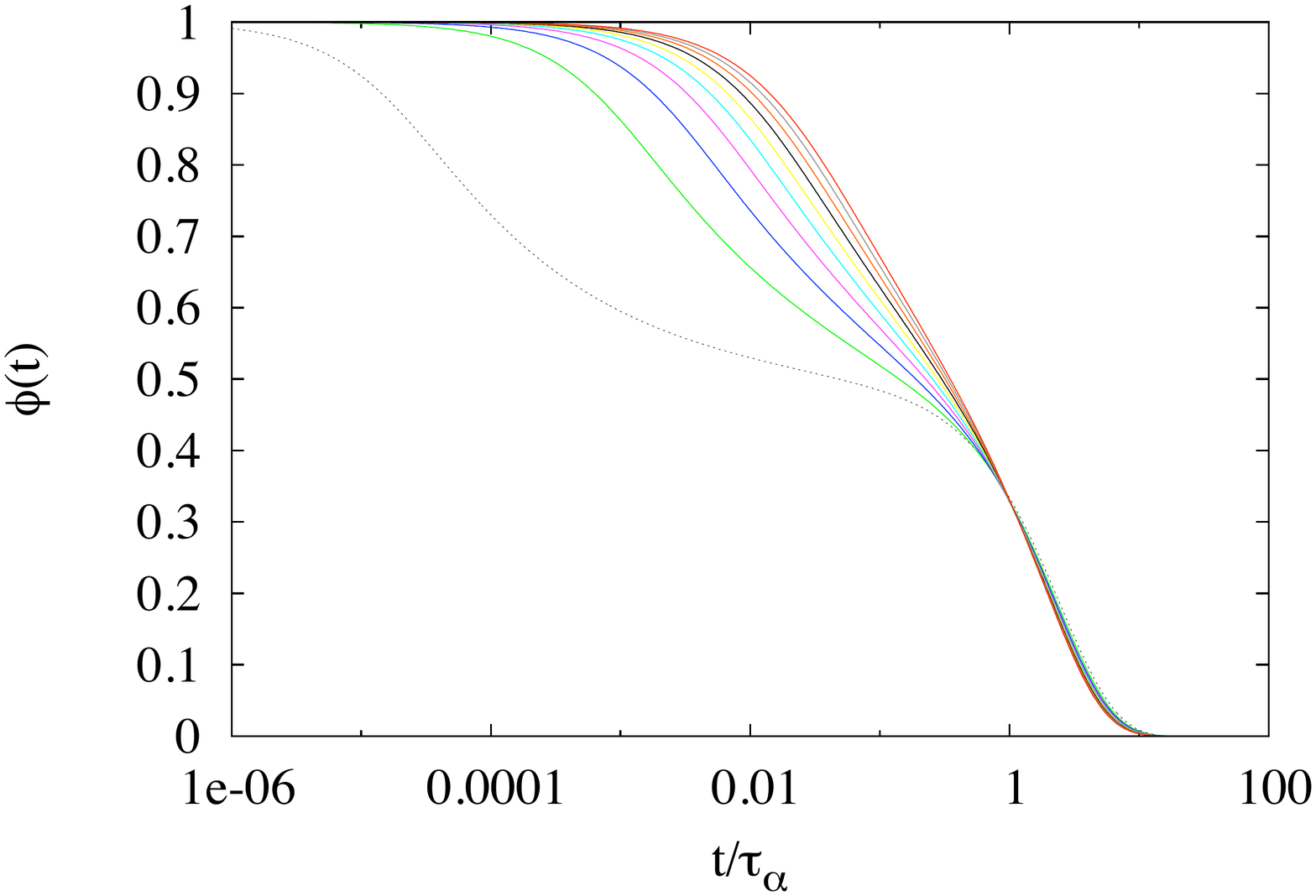}
\includegraphics[width=170pt,angle=90]{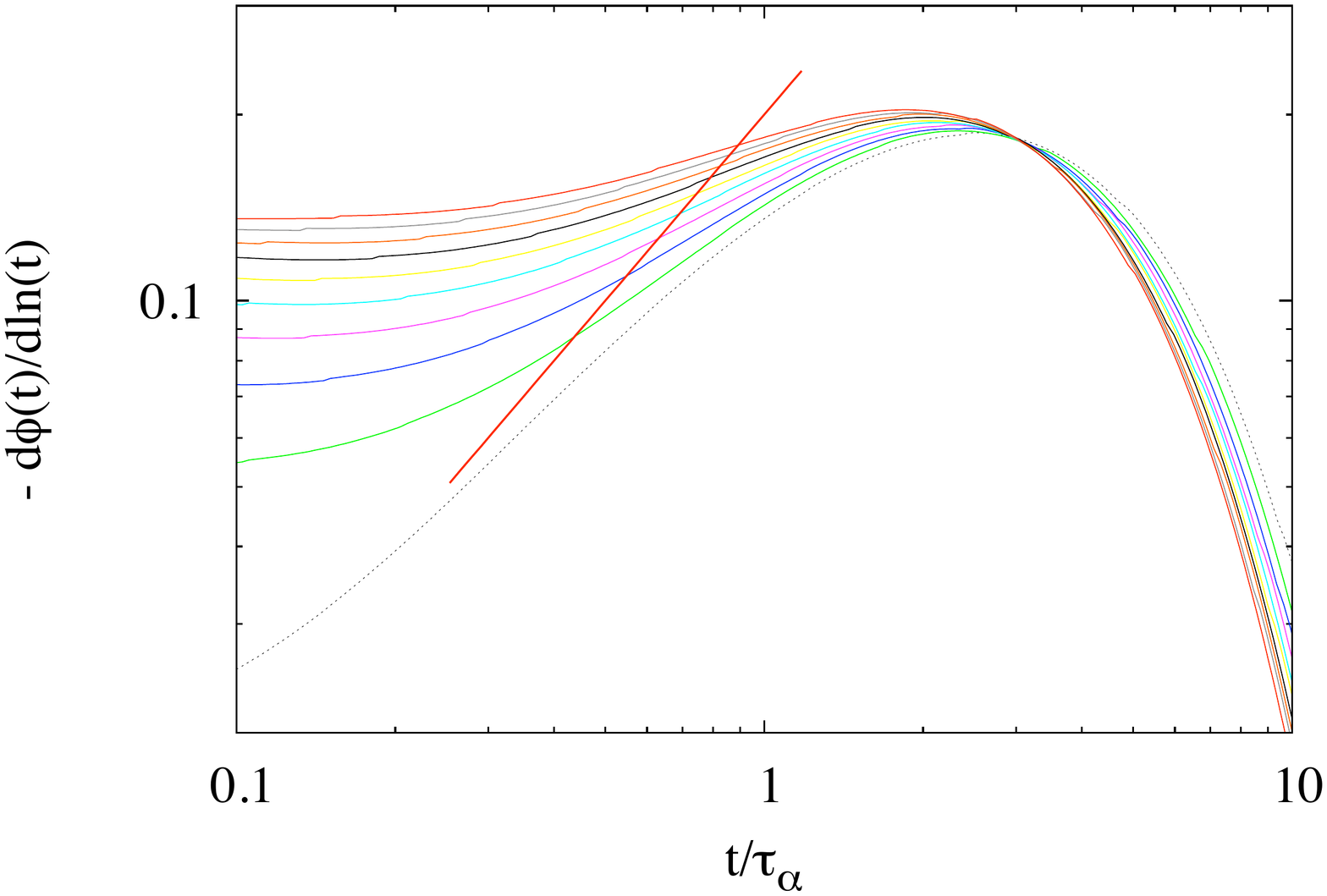}
\caption{Leutheusser model. Left: Scaling plot of $\Phi(t)$ as a
function of $t/\tau_\alpha$, at the same relative distances from the
MCT transition as for our ROM data above $T_d$, namely
$\lambda=T_d/T$, for $T=0.178, 0.186,\ldots,0.250$.  The $\alpha$ time
scale is given by $\tau_{\alpha}=(1-\lambda)^{-1.765}$. Right: Scaling
plot of $-d\Phi(t)/d\ln t$.
The expected late
$\beta$ behavior $t^b$ with $b=1$ is shown for comparison. The
straight line represents $y= t /(5 \tau_{\alpha})$.}
\label{figA2c}
\label{figA2a}
\end{figure}

Figure~\ref{figA2a} shows a TTS plot of $\Phi(t)$ using the exact
value of $\gamma$, i.e. using $\tau_{\alpha}=\epsilon^{-1.765}$, where
$\epsilon=(1-\lambda)=(T-T_d)/T_d$. This figure shows apparent scaling
in the late $\beta$ (von Schweidler) regime, with only small scaling
violation. However, only the relaxation corresponding to the value of
$\lambda$ closest to unity ($\epsilon \approx 0.05$) reveals the
expected two-step relaxation with a nontrivial plateau region!

More revealing is a log-log plot of $-d\Phi(t)/d\ln t$, shown in
figure~\ref{figA2c} together with the expected theoretical behavior
$-d\Phi(t)/d\ln t\propto t^b$ with $b=1$ (The derivative is computed
by plain finite difference). We now see very strong scaling violations
before the peak, and an apparent value of $b$ that is significantly
below $1$, even for the curve closest to the critical point. The
conclusion is that while the value of $\gamma$ extracted from a TTS
plot of $\Phi(t)$ is reasonable, the value of $b$ that one can extract
from $\Phi(t)$ $5 \%$ away from the critical point is grossly
underestimated. This is similar to our observations above for the ROM.

Let us now turn to the non linear susceptibilities $\chi_4(t)$ and
$\chi_T(t)=d\Phi(t)/d \lambda$. Within MCT, both quantities have
the same scaling behavior.  In particular one expects that
$\chi_4(t)=\epsilon^{-1}F(t/\tau_{\alpha})$, with $F(x)\propto x^b$,
in the late $\beta$ regime.  Numerically, $\chi_T(t)$ is easy to
obtain from the value of $\Phi(t)$ for different values of
$\lambda$. The case of $\chi_4(t)$ is less straightforward.  It turns
out~\cite{chi4a} that a dynamical susceptibility, proxy of $\chi_4(t)$, 
can be computed for the spherical p-spin model, which as recalled before
is characterized by a dynamical equation for the correlation function identical
to the the one of the Leutheusser model.  
A perturbed-time dependent Hamiltonian $H_{\eta}$ is
introduced:
\begin{equation}
H_{\eta}=H-\eta q_d(t)
\end{equation}
where $H$ is the usual p-spin Hamiltonian, and $q_d(t)$ is the overlap
between the spin configurations at times $t$ and $t=0$. Then
$\chi_4(t) \equiv d \langle q_d(t)\rangle/d\eta\Bigr|_{\eta=0}$.  For
a given value of $\eta$, one is led to a set of two integro
differential equations that can be solved numerically~\cite{Kim:2001rz}. The estimate of
$\chi_4(t)$ follows from a careful extrapolation to $\eta=0$.
\begin{figure}[!htb]
\centering 
\includegraphics[width=220pt,angle=0]{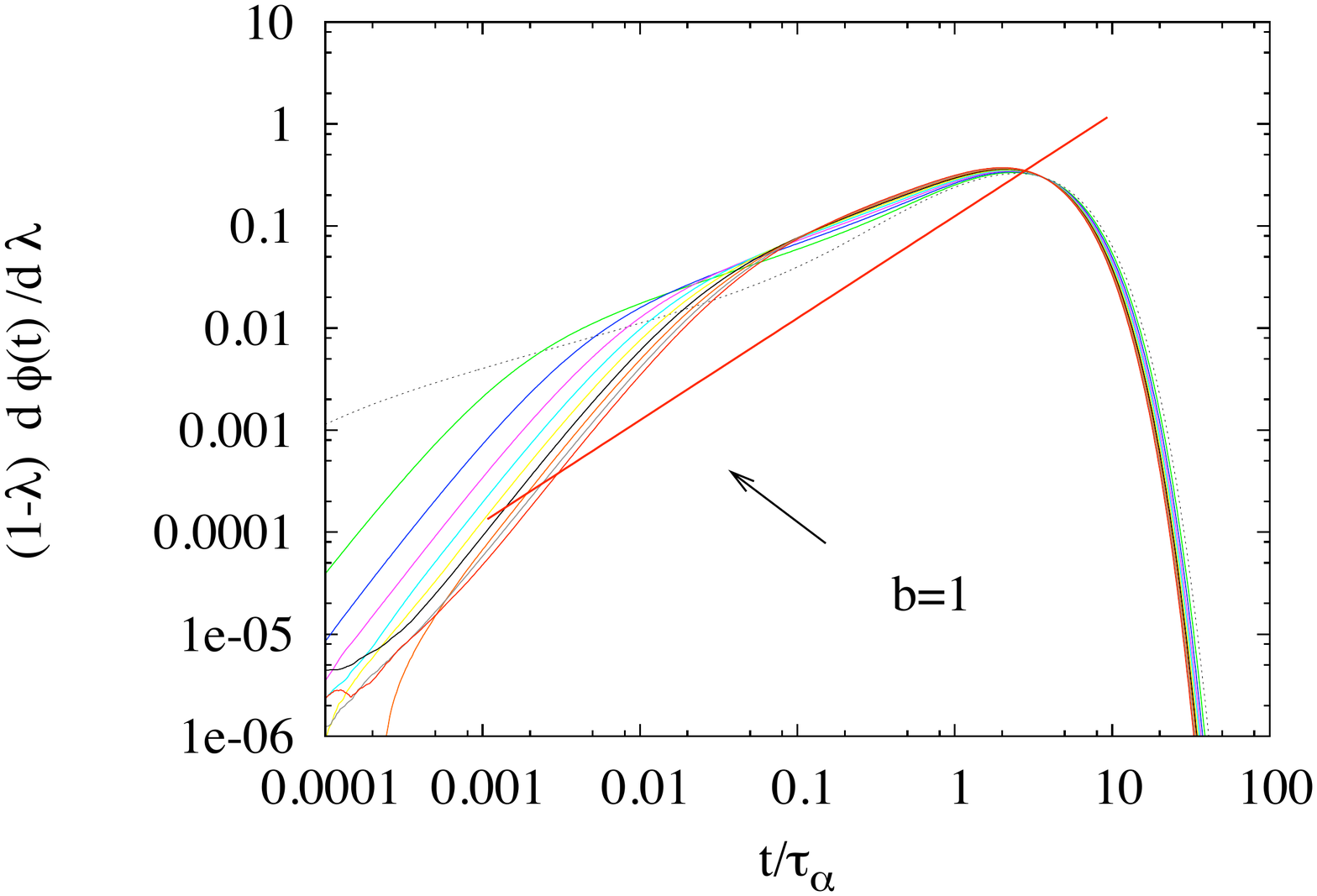}
\includegraphics[width=220pt,angle=0]{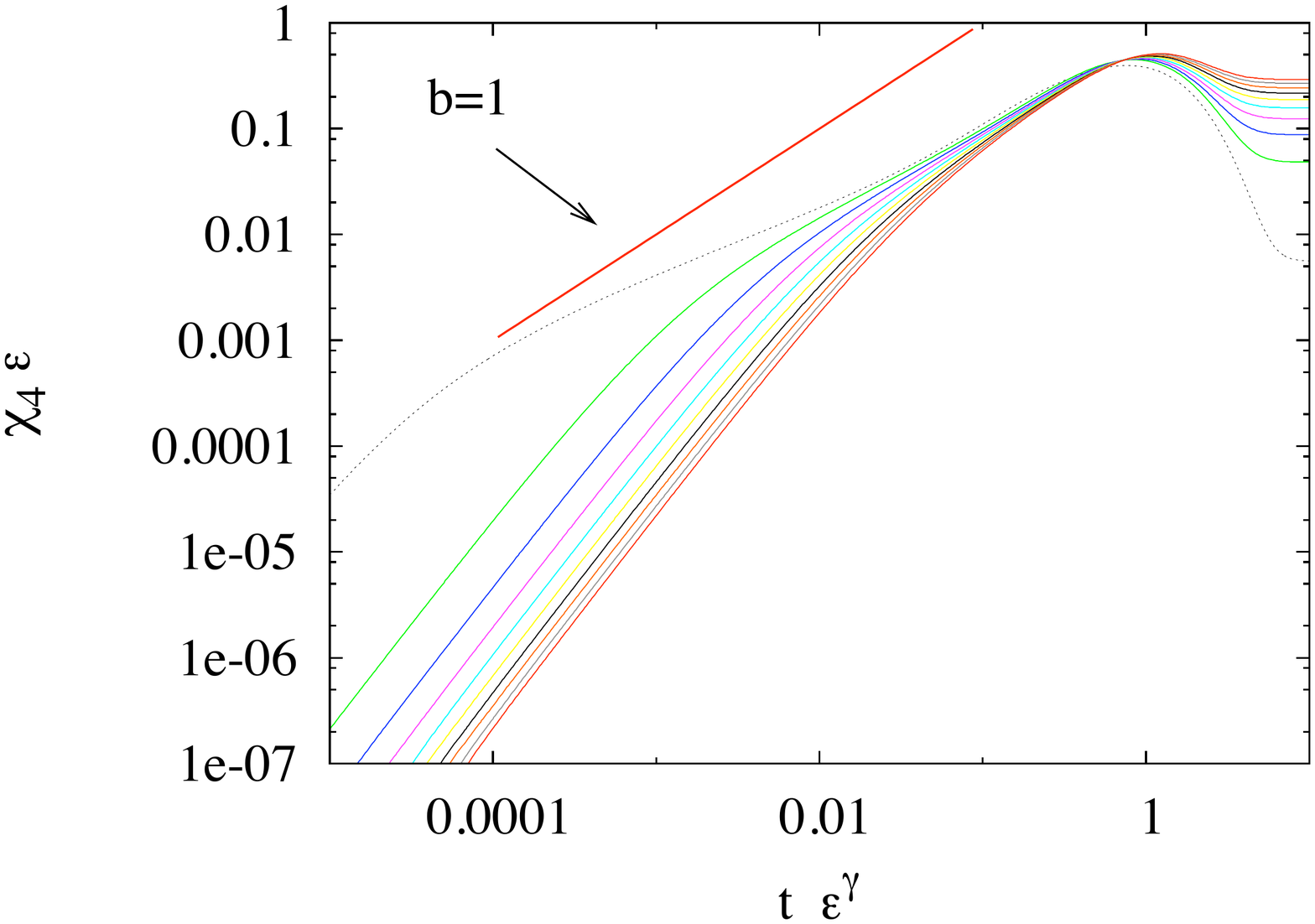} 
\caption{Left: Scaling plot of $\chi_T(t)=d\Phi(t)/d\lambda$. The
values of $\lambda$ are the same as in figures~\ref{figA2a}, with the
same color code.  The expected scaling behavior $d\Phi(t)/d\ln
\epsilon \propto (t/8 \tau_{\alpha})^b$ with $b=1$ is shown for
comparison. Right: Scaling plot of $\chi_4(t)$ for the spherical
3-spin model, with again the same values of $\epsilon$ and the same
color code.  The straight line corresponds to the theoretical
prediction $\chi_4\propto\epsilon^{-1}(t\epsilon^{\gamma})^b$.}
\label{figA2e}
\label{figA2d}
\end{figure}
Figures~\ref{figA2d} shows scaling plots for $\chi_T(t)$
and $\chi_4(t)$.  Although an approximate scaling is observed close to
the peak, only the $\chi_T(t)$ curve closest to the transition gives a
hint of the correct value of the exponent $b$.  The scaling violations
are non monotonous and can fool the reader into seeing scaling with
some $b$ less than the correct value. The estimate for $b$ from
$\chi_4(t)$ is systematically larger, and closer to the true
value. This again is similar to our numerical observations for the ROM.

\begin{figure}[htb]
\centering
\includegraphics[width=220pt,angle=0]{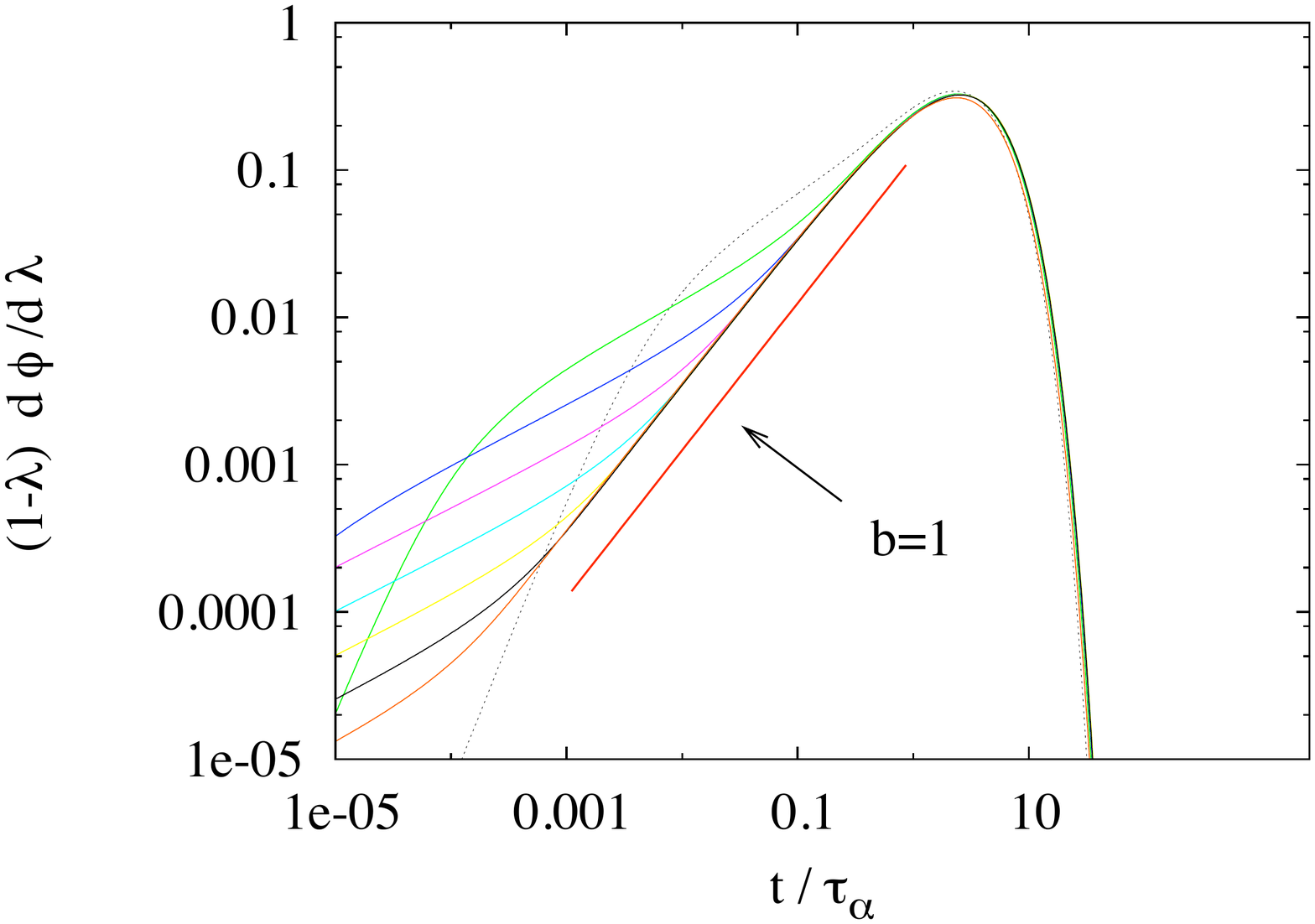}
\includegraphics[width=220pt,angle=0]{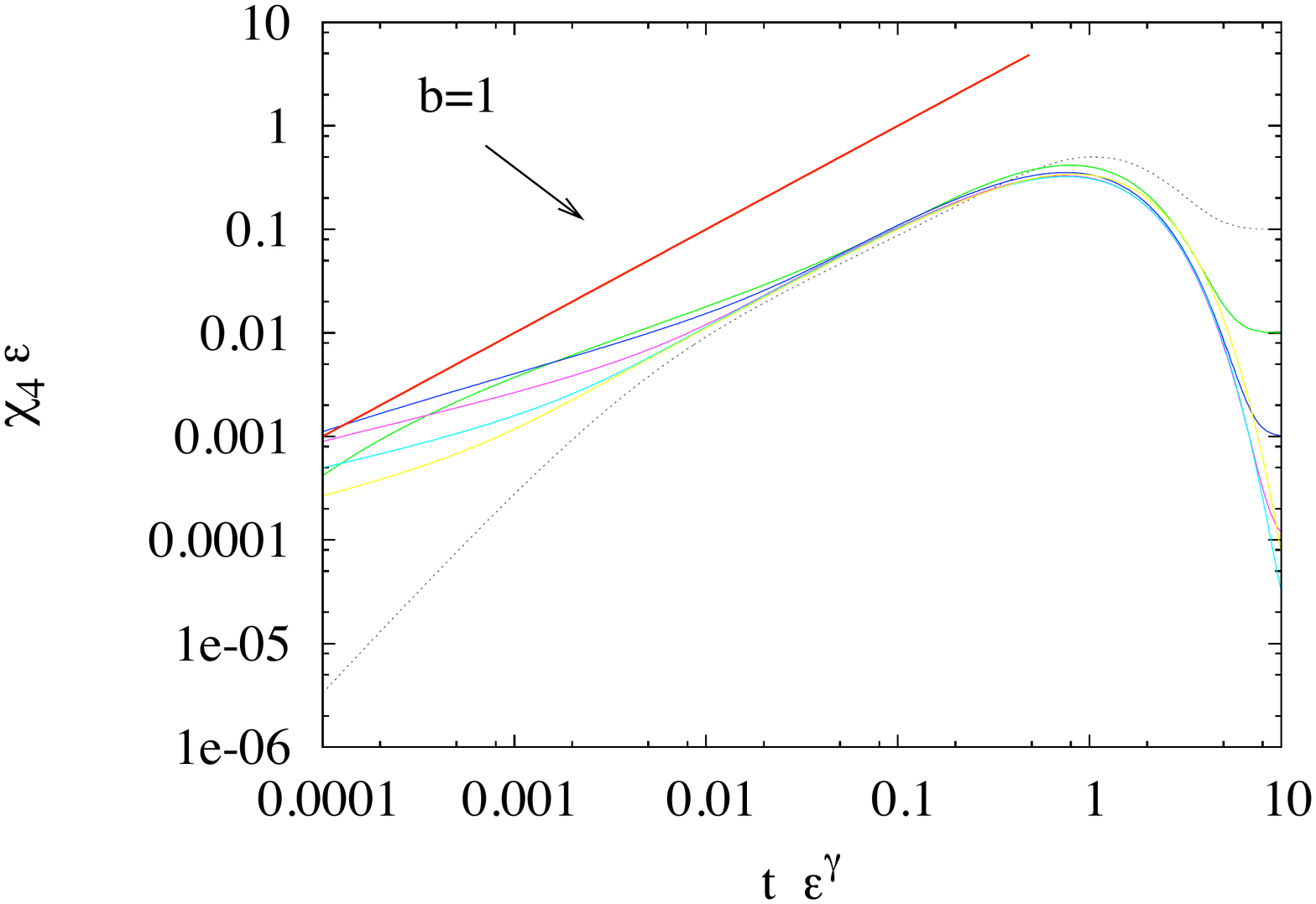}
\caption{Same as figure~\ref{figA2d}, but going deeply inside the
scaling region, namely for $\epsilon=10^{-1},10^{-2},\ldots, 10^{-8}$
for the left wing figure, $\epsilon=10^{-1},10^{-2},\ldots 10^{-6}$
for the right wing figure.  The straight line are the same as in
figure~\ref{figA2d}. }
\label{figA2h}\label{figA2i}
\end{figure}

In order to observe the asymptotic MCT scaling predictions, one must
work much closer to the transition.  For example, the expected linear
regime of $-d\Phi(t)/d\ln t\propto (t/\tau_{\alpha})$ only appears very
slowly as $\lambda\to 1$, and is well developed only for $\epsilon
\leq 10^{-4}$. Figure~\ref{figA2h} shows a similar behavior for
$d\Phi(t)/d\lambda$. The linear region appears only when
$\epsilon\leq 10^{-3}$. Finally figure~\ref{figA2i} shows the even
slower approach to scaling of $\chi_4(t)$ (beware however that
$\epsilon$ is here limited to $10^{-6}$).

\subsection{Dynamics: Conclusion}

The above results show that the true asymptotic regime of MCT is
unusually narrow. This should be remembered when comparing numerical
(or experimental) data with MCT predictions. These data are usually
plagued with noise and possible finite size effects, on top of the
strong scaling violations that appear even in the best case situation
studied in this section. On the other hand, some useful conclusions
emerge, which allows us to make sense of our data on the ROM dynamics
in the preasymptotic regime: (i) an approximate TTS holds for the late
$\beta$ regime, with the correct value of exponent $\gamma$; (ii) the
exponent $b$ extracted from the time dependence of the correlation
function underestimates the true value; (iii) for the dynamical
susceptibility $\chi_4$, the scaling is acceptable around the peak
with the predicted divergence $\epsilon^{-1}$, with a value of $b$ in
the correct range. From these considerations, we conclude that the
correct values of $b$ for the ROM should be around $b \approx 0.8$,
and the corresponding value of $\gamma$ close to $2$, and $a \approx 0.35$. A confirmation
of these values should come from studying the dynamics closer to
$T_d$. However, finite size corrections become important there and we
now turn to the study of these effects.

\section{\label{sec:4} Finite-Size Scaling: More surprises}

In the previous section, we have shown that the critical behavior of
the ROM dynamics in the regime where finite size corrections are small
is polluted by strong preasymptotic effects. In order to get rid of
those one should simulate very large systems very close to $T_d$,
which is alas not possible since the equilibration time also becomes
very large. The hope would be to use finite size scaling (FSS) to
extract the interesting asymptotic behavior.

\subsection{\label{sec:4-0}Naive theory and comparison with numerical data}

The MCT predictions are modified for
finite but large system sizes. One expects in particular that
activated effects, absent for infinitely large systems, start playing
a role for finite systems close to $T_d$.

As we have recalled above, it was recently recognized that MCT is a
mean field (Landau) theory characterized by a diverging length scale
$\xi \propto \epsilon^{-1/4}$~\cite{BB-EPL,IMCT} and the corresponding
upper critical dimension is $d_u=6$.  Assuming that the field
theoretical analysis of~\cite{Brezin:1982fj,Brezin:1985kx} applies
also to this dynamical transition
\footnote{Note that although a direct field theoretical analysis of
FSS for the MCT dynamical transition seems very difficult, the scaling
MCT exponents are related to the ones obtained from the replica
theory.  Naively, the analysis of~\cite{Brezin:1982fj,Brezin:1985kx}
is expected to hold for the replica field theory.} we expect that
finite size scaling holds for MCT above the upper critical dimension
$d_u$ where the proper scaling variable is not
$L/\xi=N^{1/d}\epsilon^\nu$ but rather
$N^{1/d_u}\epsilon^\nu$~\cite{Binder:1985kx}. The fully connected ROM
is obviously above the upper critical dimension and the relaxation time
$\tau_\alpha$ for a finite system should therefore take the following
scaling form:
\begin{eqnarray}\label{FSStau}
\tau_\alpha(T,N)&=& N^{\frac{\gamma}{\nu d_u}}
\mathcal{F}
(N^{\frac{1}{\nu d_u}} \epsilon), \qquad \nu d_u=3/2
\end{eqnarray}
When $N \to \infty$, all $N$ dependence should disappear and the MCT
divergence of equation(\ref{eq:tau}) must be recovered.  This means that the
scaling function $\mathcal{F}$ must behave as $\mathcal{F}(x)\propto
x^{-\gamma}$ when $x \gg 1$.

\begin{figure}[phtb]
\begin{center}
\includegraphics[width=220pt,angle=0]{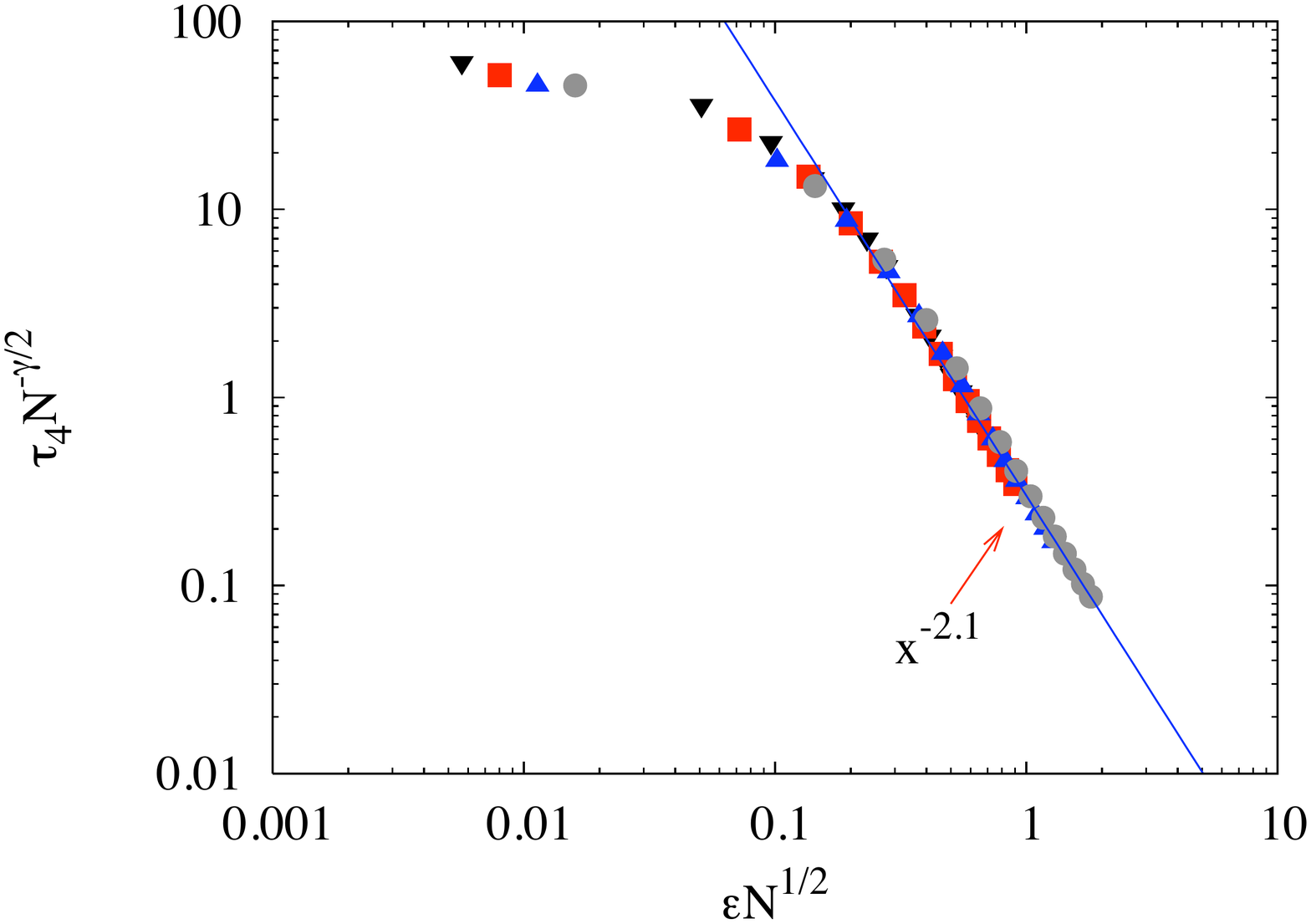}
\includegraphics[width=220pt,angle=0]{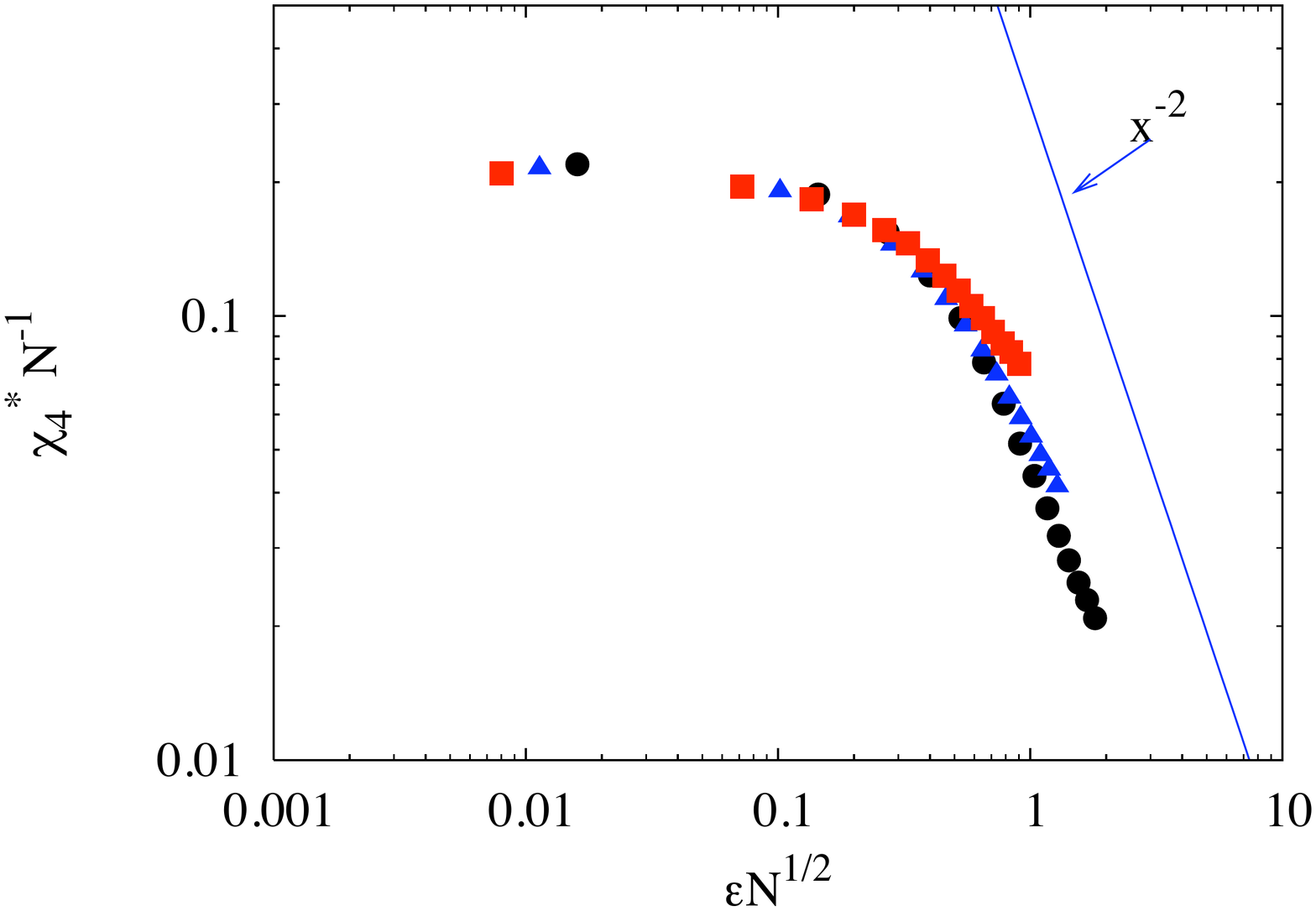}
\caption{Left: Scaling plot of $\tau_\alpha(T,N)$ using equation (\ref{FSStau}) with $\gamma=2.1$ and $d_u \nu =2$.
Right: Scaling plot of the maximum of $\chi_4$, using equation (\ref{FSSchi4}) with $d_u \nu =2$ and $s = 1$.}
\label{fig:a10}\label{fig:a14}
\end{center}
\end{figure}

We analyze our numerical results on $\tau_\alpha(T,N)$ using the above
FSS form, see Fig~\ref{fig:a10}.  A good collapse of the different
curves can indeed be obtained using equation (\ref{FSStau}) with
$\gamma=2.1$, but we need to use the value $d_u \nu =2$ instead of the expected
value $3/2$.

If we now turn to the four-point susceptibility $\chi_4$, another
confusing result is obtained. While  the peak location
$\tau_4^*$ has the same finite size scaling as  $\tau_\alpha$, as expected, the FSS of
the peak height should read:
\begin{eqnarray}\label{FSSchi4}
\chi_4^*(T,N)&=& N^{s}\mathcal{G}(N^{\frac{1}{\nu d_u}} \epsilon).
\end{eqnarray}
Using the above effective value $\nu d_u=2$, the exponent $s$ should
be such that for large $N$, $\chi_4^*$ diverges as $\epsilon^{-1}$
independently of $N$. This fixes $s=1/2$, at variance with our
numerical data that suggests $\chi_4^* \propto N^s$ with $s\simeq 1$, see
figure~\ref{fig:a14}.  Note that despite the uncertainties on the exact
value of $s$, we clearly find that $s>1/2$ (note also the decay of the
scaling function $\mathcal{G}(x)$ at large arguments).

We therefore find that finite size scaling appears to work for the ROM
transition but not in the way expected, at least naively.  In
particular, the scaling variable $\sqrt{N}\epsilon$ is at odds with
the value of the upper critical dimension for the ROM dynamics
(without explicitly conserved variables) and the value of the exponent
$\nu$.

The next sections are dedicated to a detailed discussion and
explanation of these puzzling results. As we shall show there is no
contradiction whatsoever and the origin of this strange FSS are sample
to sample fluctuations of the critical temperature $T_d$.

\subsection{\label{sec:4-1} Harris criterion and random critical points}

It is well known that the stability of pure critical points with
respect to weak disorder~\cite{Harris:1974ly}  is governed by the Harris criterion: near a
{\itshape{second order}} phase transition in dimension $d$, the bond
disorder is irrelevant if the specific heat exponent $\alpha_{pure} =
2 - d\nu_{pure}$ is negative, where $\nu_{pure}$ is the correlation
length exponent of the pure system.  Then the critical exponents of
the disordered system are the same as the ones of the pure system.

If $\alpha_{pure} >0$, disorder becomes relevant and the system is
driven towards a so called random fixed point characterized by a new correlation
length exponent $\nu_R$ satisfying the general bound $\nu_R\ge
2/d$. In the last twenty years, important
progress~\cite{Chayes:1986kx,Aharony:1996ys,Wiseman:1998vn} has been
made in the understanding of finite size properties of random critical
points.  For our purpose we only recall that to each
realization of disorder ($\omega$), one can associate a
pseudo-critical temperature $T_d(\omega,N)$, defined for instance as
the temperature where the relevant susceptibility is maximum. The
disorder averaged pseudo-critical critical temperature
$T^{N}_d\equiv\overline{T_d(\omega,N)}$ converges towards its infinite
size limit as: $T^{\infty}_d-T_d^N \propto N^{-1/d\nu_R}$. The 
width $\Delta T_d(N)$ of the
distribution of the pseudo-critical temperatures $T_d(\omega,N)$
then  depends on the nature of the critical point. If
the disorder is irrelevant, $\Delta T_d(N)$ scales trivially like
$N^{-1/2}$, but like $N^{-1/d\nu_R}$ if the disorder is relevant. 

\begin{figure}[phtb]
\begin{center}
\includegraphics[width=200pt,angle=0]{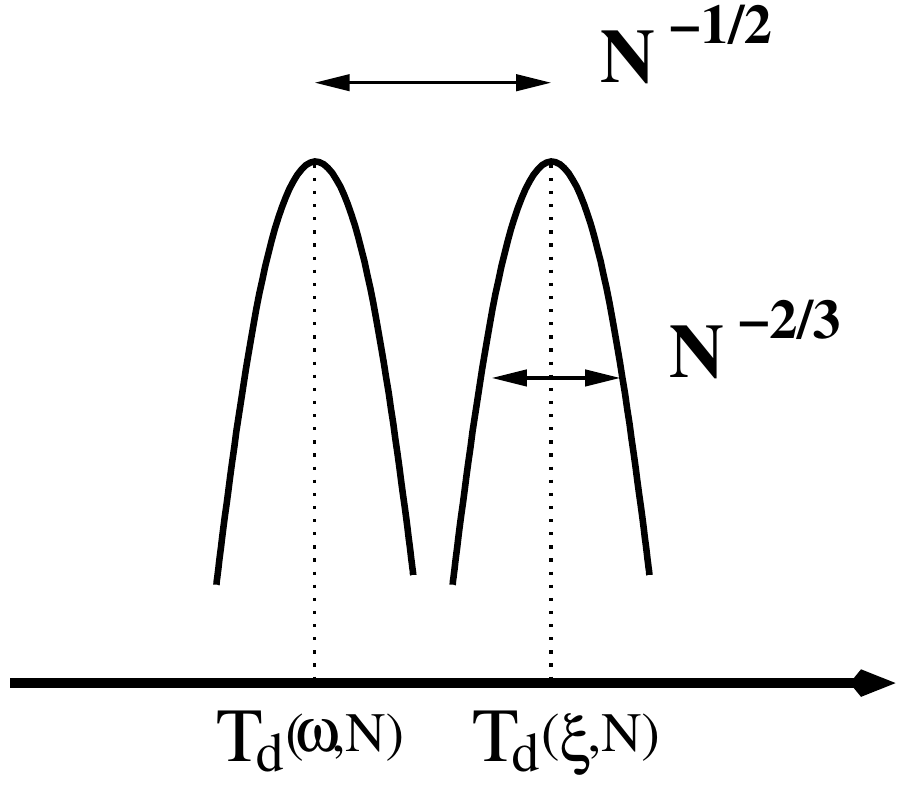}
\caption{Sketch of the sample to sample fluctuations of a
susceptibility in situations where the critical temperature fluctuates
on scales $N^{-1/2}$, much larger than the FSS thermal window
$N^{-2/3}$.  Here $\omega$ and $\xi$ are two disorder configurations.
}\label{fig:051}
 \end{center}
\end{figure}

All previous studies of FSS for disordered systems mentioned above 
have focused on models below their upper critical dimension (see for
example~\cite{GarelMonthus}). Instead the ROM is clearly above its
upper critical dimension.  As a consequence, it is not obvious to
deduce its FSS behavior from previous works. In~\ref{sec:A-1}
we discuss in detail the subtleties of the Harris criterion above the
upper critical dimension.

From a phenomenological point of view, one expects that general
features of random fixed points should still occur: in particular one
can define a sample-dependent pseudo-critical dynamical temperature
$T_d(\omega,N)\equiv T_d+\delta T(\omega,N)$.  A hand-waving argument
to understand the origin of these fluctuations is to consider the TAP
equations for the ROM. The high temperature expansion leading to the
TAP equations for the ROM has disorder-dependent corrections of order
$N^{-1/2}$.  These have a dramatic effect on FSS since these
corrections are much larger than the expected FSS thermal window
$N^{-2/3}$ of the MCT-transition. We will therefore assume, and
justify later on, that for each sample the FSS window is indeed of the
order $N^{-2/3}$ around a random critical temperature that has
disorder fluctuations of the order $N^{-1/2}$. As a consequence, FSS
for disordered averaged observables are dominated by the fluctuations
of the critical temperature that wash out the much sharper $N^{-2/3}$
FSS thermal window~\footnote{This cannot 
happen below the upper   critical dimension due to the Harris criterion.} (see figure~\ref{fig:051} for a cartoon
representation). 
Note that a similar situation for FSS of random first
order transition has been discussed by D.S. Fisher
in~\cite{Fisher}.

\subsection{\label{sec:4-2} Random critical temperatures and modified FSS}

To make the above statements more precise, let us consider the generic
example of some thermodynamic observable $\mathcal{O}$. One would like
to compute $\overline{\mathcal{O}}$. We assume that averaging over the
disorder is equivalent to averaging over the distribution of critical
temperatures, $p(T_d(\omega,N))$. For the sake of simplicity,
$p(T_d(\omega,N))$ is taken to be a well behaved distribution with
width of order $N^{-1/2}$ centered on $T_d$, the true asymptotic
dynamical temperature \footnote{There will also be finite size corrections 
to the center of the distribution, but these are expected to be subleading to $N^{-1/2}$.}. Moreover, we posit that for each sample, some
kind of FSS holds, in the sense that:
\begin{eqnarray}
\mathcal{O}(T,N)=N^{2 \zeta/3} \mathcal{F}(N^{2/3}[T-T_d(\omega,N)]) 
\label{eq:fss}
\end{eqnarray}
where $\zeta$ is a certain exponent that depends on the particular
observable, and the scaling function $\mathcal{F}$ is, at least to
leading order, sample-independent. We assume that $\mathcal{F}(x)$ is
regular for small arguments, and $\mathcal{F}(|x|) \propto A_\pm
|x|^{-\zeta}$ for $x \to \pm \infty$, such that $\mathcal{O}(T,N \to
\infty) \propto |\epsilon|^{-\zeta}$ independently of $N$. It is
possible to justify all these assumptions within a simple toy model,
the weakly disordered version of the Blume-Capel model, that displays
exactly the unusual FSS discussed in this section. We warmly invite
the reader to examine~\ref{sec:A-2} for more details.

Now, writing $T_d(\omega,N) = T_d - y N^{-1/2}$, disorder averaging is
obtained by computing:
\begin{eqnarray}
\overline{\mathcal{O}}(T)&=&N^{2\zeta/3} \int dy \, p(y) \mathcal{F}(N^{2/3}\epsilon+ N^{1/6}y) 
\end{eqnarray}
where $\epsilon=T-T_d$. The analysis of the above integral in the large $N$ limit requires to distinguish two cases: $\zeta < 1$ and $\zeta > 1$ 
(with further logarithmic terms when $\zeta=1$). 
\begin{itemize}
\item When $\zeta < 1$, one can set $\epsilon=u/\sqrt{N}$ and take $N
\gg 1$ in the above integral, to get to leading order:
\begin{eqnarray}
\overline{\mathcal{O}}(T)&=& N^{\zeta/2}
\hat{\mathcal{F}}(\sqrt{N}[T-T_d]),\\
\hat{\mathcal{F}}(u)&=&\int_{-u}^\infty dy \,
\frac{A_+p(y)}{(u+y)^\zeta} +\int_{-\infty}^{-u} dy \,
\frac{A_-p(y)}{|u+y|^\zeta}
\end{eqnarray}
Since $\zeta < 1$, the integral defining $\hat{\mathcal{F}}$ is always
convergent when $u \to 0$, leading to a well behaved scaling
function. Therefore, in this case the usual FSS strategy is valid,
with a scaling variable dominated by the fluctuations of critical
temperature: $N^{2/3} \to N^{1/2}$.
\item When $\zeta > 1$, on the other hand, the relevant change of variable is $y=z/N^{1/6}-u$. Again to leading order, this gives:
\begin{eqnarray}
\overline{\mathcal{O}}(T)&\approx& A N^{\frac{4\zeta-1}{6}}
p\left(-\sqrt{N}[T-T_d]\right)
\end{eqnarray}
where $A=\int dz \, \mathcal{F}(z)$ is a convergent integral thanks to
the rapid decay of $\mathcal{F}(z)$ for large $z$, leading to a
finite multiplicative constant. In this case, FSS is drastically altered by
the sample to sample fluctuations of the dynamical temperature; the
decay of the scaling function is related to the one of the
distribution of critical temperatures, and the exponent
${(4\zeta-1)}/{6}$ is unusual.
\end{itemize}
The above analysis can be extended to the case of ``assymetric
observables'', where the power-law decay of $\mathcal{F}(x)$ is
different when $x \to +\infty$ and $x \to -\infty$. Most of the
interesting ROM observables turn out to be of that type around $T_d$,
see section~\ref{sec:4-4},~\ref{sec:4-5}.

\subsection{\label{sec:4-3} FSS for single samples}
\begin{figure}[phtb]
\begin{center}
\includegraphics[width=220pt,angle=0]{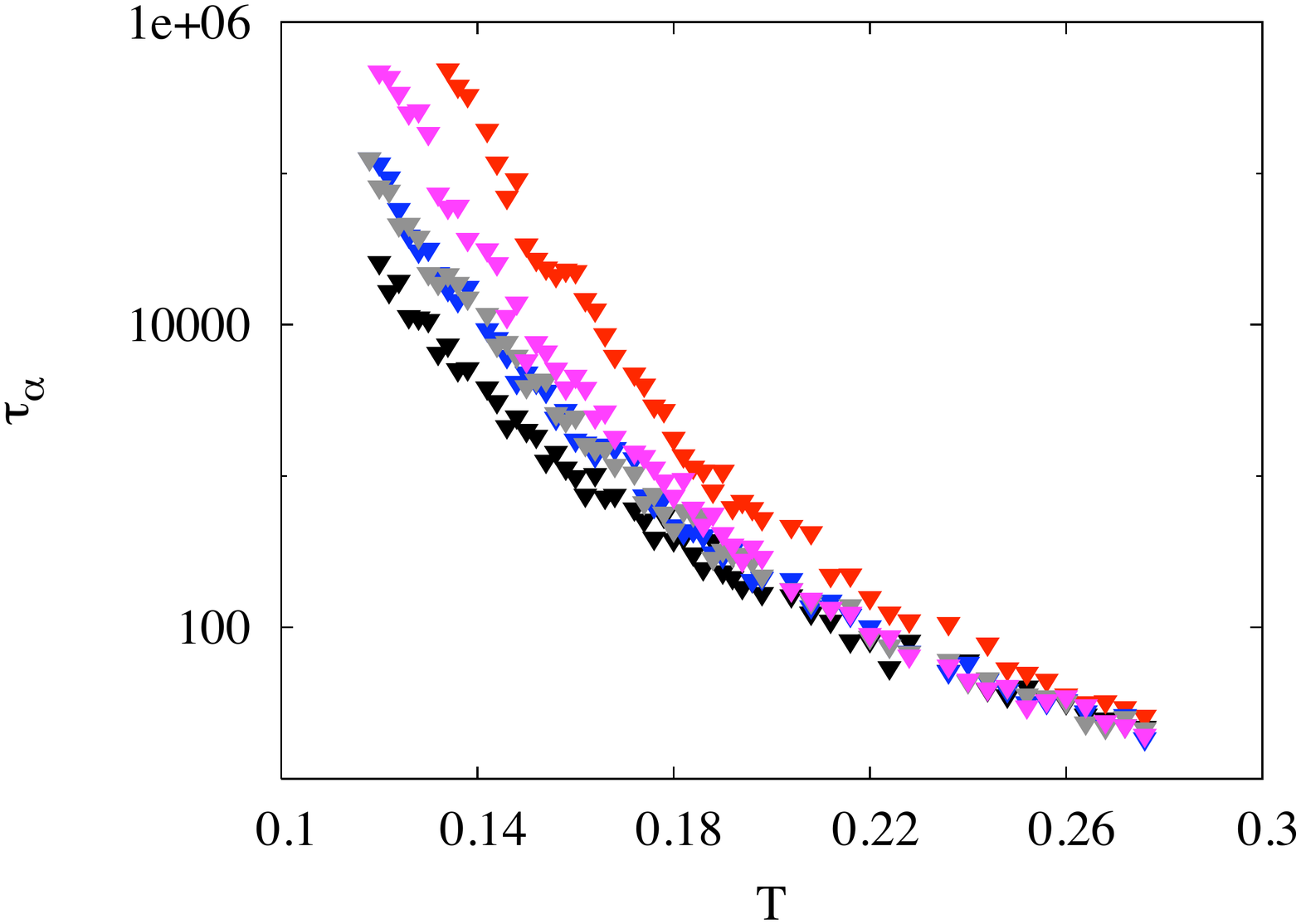}
\includegraphics[width=220pt,angle=0]{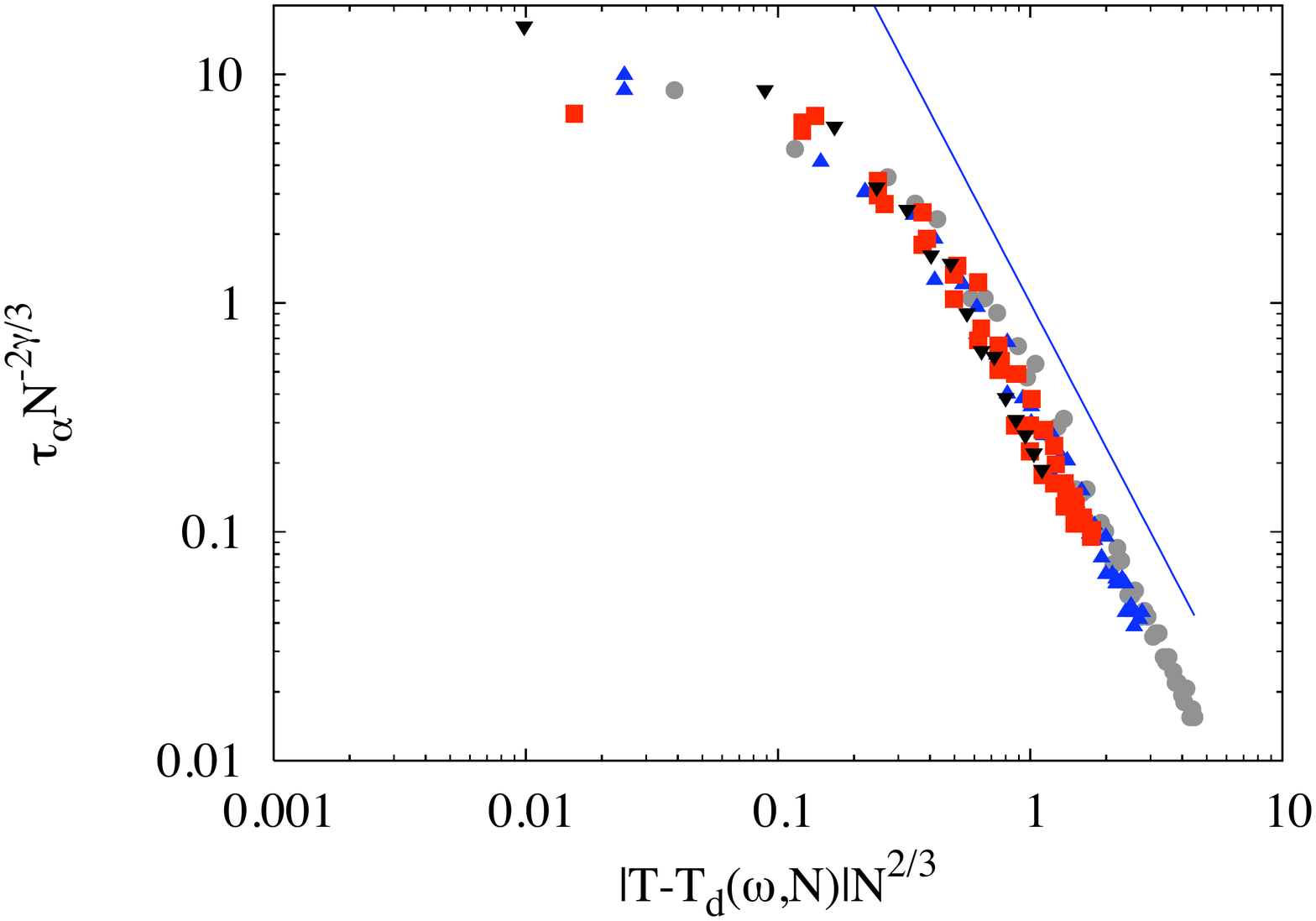}
\caption{Left: $\tau_{\alpha}$ is plotted as a function of the
  temperature for 5 different disorder samples ($N=32$ with $N_{ther}=120$ only). There are
  clear strong sample to sample fluctuations of the relaxation
  time. Close to $T_d$, the five curves can be superposed by a
  suitable horizontal shift, namely one has the relation
  $\tau(\omega,N)=\mathcal{C}(T-T_d(\omega,N))$.  Right: Scaling plot
  of the relaxation time $\tau_{\alpha}$ using rescaled sample dependent
  dynamic temperature.  Here are plotted the fastest, the slowest and
  a typical sample for all sizes (N=256 corresponds to grey circle,
  N=128 to upper blue triangle, N=64 to red squares and N=32 to lower
  black triangle). We recovers the standard FSS exponents by using
  $T_d(\omega,N)$ instead of the thermodynamic $T_d$} 
  \label{fig:026}
\end{center}
\end{figure}

Checking these assumptions numerically is tricky for the ROM. If there
was a quantity (susceptibility-like) with a sharp peak around $T_d$
then it would be simple: one would just rescale for each sample this
quantity around its peak and verify whether the usual finite size
scaling holds, as suggested by figure~\ref{fig:051}.  Unfortunately,
no such quantity exists for the ROM. As we have seen, $\chi_4^*(T)$ is
a monotonously decreasing function of temperature. In order to check
the usual FSS, one should shift horizontally $\chi_4^*(T)$ for each
samples around its own effective dynamical transition temperatures
$T_d(\omega,N)$. In order to determine this effective critical
temperature we focus on the sample to sample fluctuations of the
relaxation time (see left panel of figure~\ref{fig:026}). We assume that the
relaxation time is uniquely determined ---at fixed $N$--- by the distance
$T-T_d(\omega,N)$, i.e. 
$\tau(\omega,T)=\mathcal{C}(T-T_d(\omega,N))$, where $\mathcal{C}$ is a certain 
function.  Thus, choosing a certain reference
relaxation time $\tau^*$ for a given sample, one fixes
the difference $T^*(\omega,N)-T_d(\omega,N)$ to a sample independent
value $\delta^*$ (with $\tau(\omega,T^*(\omega,N))=\tau^*$)
. 
Therefore, by averaging,
$$ \delta^*=T^*(\omega,N)-T_d(\omega,N) =
\overline{T^*(\omega,N)-T_d(\omega,N)} = \overline{T^*(\omega,N)}-T_d
+ O(N^{-2/3}),
$$ where we have assumed that the $N$ dependent correction to the
average critical temperature is $O(N^{-2/3})$, that only introduces a
shift in the final scaling variable. The above equation allows one to
determine the sample dependent shift of critical temperature,
$T_d(\omega,N)-T_d$, as
${T^*(\omega,N)}-\overline{T^*(\omega,N)}$. This procedure does not
require the knowledge of the functional form of the relaxation
time. Once this shift is know, one can rescale the temperature axis in
a sample dependent way, and test FSS sample by sample.

\begin{figure}[phtb]
\begin{center}
\includegraphics[width=300pt,angle=0]{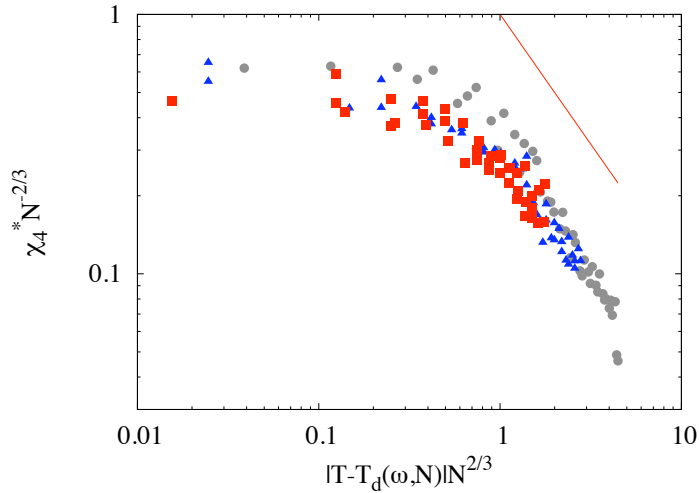}
\caption{Scaling plot of the maximum of $\chi_4$ with sample dependent transition
  temperatures $T_d$ (same convention as figure~\ref{fig:026}). The behavior
  of $\chi_4^*(T)$ is now compatible with the MCT-predictions,
  $\chi_4^*(T) \propto \epsilon^{-1}$ and with the standard
  finite-size scaling exponents.}\label{fig:031}
 \end{center}
\end{figure}
The results are given in figures~\ref{fig:026} and~\ref{fig:031}. Our
statistics is quite limited because of the CPU-time consumption of
such simulations. The modest number of thermal configuration does not
allow us to extract the whole probability distribution for
$T_d(\omega,N)$. Despite these limitations, the results are in perfect
agreement with our expectations. After the above temperature
rescaling, the best collapse of the relaxation time data is obtained
with the naive finite-size scaling ($N^{2/3}(T-T_d(\omega,N)))$ and not
$\sqrt{N}(T-T_d(\omega,N))$. A similar behavior is obtained for
$\chi_4^*$. The dynamical susceptibility divergence is now compatible
with an $\epsilon^{-1}$ behavior derived analytically for MCT
transition.

\subsection{\label{sec:4-4} Anomalous FSS for the dynamical susceptibility}

Now we are confident in the validity of our analysis in terms of
sample dependent transition temperatures, we come back on the
anomalous FSS properties of the dynamical susceptibility that we
observed numerically in section~\ref{sec:4-0} above. We must first guess
the shape of the sample dependent FSS for the peak
susceptibility. When $T > T_d(\omega,N)$, we expect a divergence of
$\chi_4^*(T)$ as $\epsilon^{-1}$, as discussed in the section
above. So the exponent $\zeta_+$ corresponding to this region is
$\zeta_+=1$. On the other hand, below the transition one expects that
the variance of the dynamical overlap will be of order $N$, due to
activated dynamics that makes the system hop between states with zero mutual overlap (see
the above remark on the fast relaxation of $q_d(t)$ in the ROM). 
Matching the requirement $\chi_4^*(T) \propto N$
with the finite-size scaling form valid near $T_d(\omega,N)$,
$\chi_4^*(T) \approx N^{2/3} \mathcal{F}(N^{2/3}(T-T_d(\omega,N)))$
leads to $\mathcal{F}(x) \sim \sqrt{-x}$ for $x \to -\infty$, or
$\zeta_-=-1/2$. Extending the analysis of section~\ref{sec:4-2} to this
strongly asymmetric case where $\zeta_+ \neq \zeta_-$, we find that
the average behavior is dominated by the left tail of
$\mathcal{F}(x)$, finally leading to:
\begin{eqnarray}
\chi_4^*(T) &\sim& N^{s} \mathcal{G}(\sqrt{N}(T-T_d)),
\end{eqnarray}
where the anomalous exponent $s$ is equal to $3/4$ and $\mathcal{G}$
is a certain scaling function. We therefore qualitatively understand
the anomalous FSS result obtained in section~\ref{sec:4-0}, in particular
the fact that $s$ is larger than the naive value $s=1/2$.

\subsection{\label{sec:4-5} Relaxation time: some conjectures}

We finally turn to the relaxation time $\tau_\alpha$, which, as we
already know, is very strongly sample dependent. The guess for the
sample dependent FSS of the relaxation time must now account for the
$\epsilon^{-\gamma}$ divergence for $T > T_d(\omega,N)$, and the
activated dynamics for $T < T_d(\omega,N)$. If we assume that the
relevant energy barrier scales with the system size as $N^\psi$, one
is led to the following ansatz:
\begin{eqnarray}
\tau(\omega,N)&=&N^{\frac{2\gamma}{3}}
 \mathcal{F}((T-T_d(\omega,N))N^{\frac{2}{3}}), \mbox{with} \\ & &
 \mathcal{F}(x)\propto \exp(-C(\omega) |x|^{\frac{3\psi}{2}}) \ \ \ x\to
 -\infty \\ & & \mathcal{F}(x)\propto x^{-\gamma} \ \ \ x\to \infty,
\end{eqnarray} 
where $C(\omega)$ is a sample dependent constant that accounts for a
possible sample dependence of the scaled barrier height. In the case
of the ROM, it is reasonable to expect 
that $\psi=1$~\cite{Ioffe:1998df,Lopatin:2000br,Lopatin:1999pz,BIL}.  The
above equation suggests that any observable governed by low enough
moments of $\tau(\omega,N)$ will correspond to negative values of the
exponent $\zeta$ in the analysis of section~\ref{sec:4-2} above. Since
our definition of the average relaxation time $\tau_\alpha$ is such
that $\overline{q_d(t=\tau_\alpha,T)}=1/2$, this quantity is dominated
by typical samples and we expect standard FSS with scaling variable
$\sqrt{N}\epsilon$, as indeed found in in section~\ref{sec:4-0}.

We however expect very different results for quantities sensitive to
large relaxation times, dominated by rare samples. For example, the
long time asymptotics of $\overline{q_d(t,T)}$ is dominated by
particularly ``cold'' samples. Neglecting the fluctuations of
$C(\omega)$, and assuming a Gaussian distribution of critical
temperatures, we find:
\begin{eqnarray}
\overline{q_d(t,T)} \sim_{t \to \infty} \int_u^{\infty}
\frac{dy}{\sqrt{2\pi \sigma^2}} \exp\left[
-tN^{-\frac{2\gamma}{3}}e^{-CN^{\psi-1/2}(y-u)^{\frac{3\psi}{2}}}\right]
e^{-\frac{y^2}{2\sigma^2}} ,
\end{eqnarray}
with $u = \sqrt{N} \epsilon$. Evaluating this integral by steepest
descent, one finds that the asymptotic relaxation regime is, to
leading logarithmic order:
\begin{eqnarray}
\ln \overline{q_d(t,T)} \propto_{t \to \infty} \left[\frac{\ln t}{N^{\psi-\frac12}}\right]^{\frac{4}{3\psi}} \qquad (\psi > \frac12).
\end{eqnarray}
This decay is far slower than a stretched exponential, and gives a
rationale to explain the observed slowing down of the late relaxation
of $q_d$, and is consistent with the behavior seen 
in the figures~\ref{fig:4A1}. In particular, we expect this
slowing down due to cold samples to become dominant close to $T_d$.

\section{\label{sec:7} Conclusions}

The aim of this study was to test numerically the predictions of MCT
in a best case situation, namely for a model with an exact MCT
transition, and analyze its finite size scaling behavior. \\
 We chose the fully connected Random Orthogonal model, with
a choice of parameter such that the dynamic (MCT) transition
temperature $T_d$ is well separated from the static (Kauzmann)
transition. We have first compared the theoretical predictions for the
static (thermodynamic) properties of the model with our numerical
results. Although we are not able to equilibrate large systems below
$T_d$, we find a good overall agreement. The transition temperature
for the ROM is very low compared to the scale of the interactions;
this implies that the transition is very strongly discontinuous, with
an Edwards-Anderson order parameter very close to unity as soon as $T
< T_d$.

We have then studied the equilibrium dynamics of the model, focusing
on the time correlation function and the four-point dynamical
susceptibility, which measures the strength of dynamical
heterogeneities. When comparing our numerical results to the
predictions of MCT, we find that while some of these predictions are
quantitatively obeyed (like approximate Time Temperature Superposition
in the late $\beta$ regime), other important ones are clearly
violated, with inconsistent values of the MCT exponents. 
This is due to strong pre-asymptotic effects. Indeed, we have shown
that the asymptotic MCT predictions are only valid inside an unusually narrow
sliver around $T_d$, thereby explaining these quantitative
discrepancies and allowing one to get rough estimates of the MCT
exponents for the ROM: $b \approx 0.8$ and $\gamma \approx 2$. Working
closer to $T_d$ to get rid of these strong preasymptotic corrections
is hampered by equally strong finite size corrections. On that front,
more surprises emerge: we find that the usual Finite Size Scaling (FSS)
fails to account for our data, a result that we rationalize in terms
of strong sample to sample fluctuations of the critical
temperature. We have developed a phenomenological theory for FSS in
the presence of these strong fluctuations. This modified form of FSS
accounts well for our results; we also show that naive FSS works for
individual samples. En passant, we have also developed new
arguments to understand FSS in disordered systems above their upper
critical dimension (see Appendices).

The compatibility between the MCT predictions and the numerical
behavior of a model that is in principle exactly described by MCT
turned out to be extremely difficult to establish quantitatively,
partly because of the impossibility to equilibrate large systems. The
situation is expected to be worse when dealing with experimental data
for which the critical temperature is blurred by non-mean field
effects. In this case, quantitative comparison with MCT requires
extreme care, to say the least. Our results show that some predictions 
appear to be more robust than others and this provides some guidance
when dealing with application of MCT to experimental or numerical data. 
Indeed, we notice that the kind of violations of MCT predictions found in our study resemble very much what found in real liquids, 
see e.g. the difference between 
the time evolution of $\chi_4$ and $\chi_T$ in \cite{BBBKKR2}.  

On a different front, our results maybe relevant
for FSS studies of super-cooled liquids~\cite{LudoFSS,IndiansPNAS}.
In this case it has been shown that the usual theory valid for 
pure systems fails in account the finite size 
scaling behavior~\cite{IndiansPNAS}. 
Although several justification can be put forward, in particular that
the correlation length is not much larger than the microscopic length, 
our results suggest that new phenomena might be at play. In fact, if the dynamically 
self-induced disorder present in super-cooled liquids plays somehow the role
of the quenched disorder present for the ROM, as often proposed, then strong disorder 
fluctuations could lead to violation of usual FSS. Results qualitatively similar
to the one reported in \cite{IndiansPNAS} (such as e.g. figure~\ref{fig:004}) are indeed
expected for the ROM. It would be certainly worth to pursuing further the 
comparative study of FSS in the ROM and real liquids.    

Finally, the study of a
finite-range ROM that replicates the phenomenology of finite
dimensional supercooled liquids is an interesting project that we are
currently pursuing, in particular to test the predictions of the
Random First Order Theory on a physical model that is as close as
possible to its theoretical idealization.  Other directions worth
investigating numerically include a better understanding of the low
temperature activated dynamics, which is probably only accessible in
the aging regime.

\subsection{Acknowledgments}
We thank A. Aharony, D.S. Fisher, T. Garel, A.B. Harris and C. Monthus
for helpful 
comments on finite size scaling properties of disordered systems and C.Dasgupta, D. Reichman
and S. Sastry for discussions on finite size scaling for super-cooled liquids. 
We thank C. Alba-Simionesco, A. Lef\`evre for discussions and A. Crisanti for 
collaboration on a related work. The numerical integration of the schematic 
MCT equations has been performed using a numerical code developed by 
K. Miyazaki whom we thank very much. Finally, we warmly thank A. Lef\`evre for a careful reading of the
manuscripts and useful comments. We acknowledge partial financial support from ANR DYNHET. 

\appendix

\section{\label{sec:A-1}FSS and Harris criterion for disordered systems above 
their upper critical dimension}

In the following we shall consider the role of disorder on finite size
scaling above the upper critical dimension. This will lead us to
formulate two different Harris criteria: one valid for the FSS and the
other for the critical exponents, obtained from susceptibilities in the
thermodynamic limit.

Let us consider a pure system that is perturbed by the addition of
small quenched disorder. As usual, we will focus on a disorder that
couples to the energy, e.g. random couplings.  As a consequence,
samples of size $N$ will be characterized by fluctuations of the
critical temperature $\delta T_c(N)$.
 
In order to understand whether disorder affects the FSS behavior one
has to compare the above fluctuations with the FSS window. Above the
the upper critical dimension FSS is subtle \cite{CardybookFSS}: the
scaling variable is $\epsilon N^{1/(d_u\nu_{u})}$ where $\nu_{u}$ is
the mean field exponent, $\epsilon=(T-T_c)/T_c$ and $d_u$ is the upper
critical dimension, see \cite{Jones:2005uq} for a numerical check in five
dimension for the Ising model.  This means that properties of a pure
finite systems depart from the ones expected in the thermodynamic
limit when the distance from the critical temperature becomes smaller
than $N^{-1/(d_u\nu_{u})}$. 
Assuming that on scales such that $\epsilon N^{-1/d_u\nu_u} \propto O(1)$
the fluctuations of the critical temperature are of the order $1/\sqrt{N}$, one finds that
for $d_u\nu_{u}<2$ and close enough to the
critical point, the addition of a small quenched disorder will
therefore make fluctuates $T_c$ on a scale much larger than the FSS
window of the pure system implying that the FSS behavior will be
drastically affected by adding an infinitesimal disorder.

This provides a generalization of the Harris criterion for FSS
properties of systems above their upper critical dimension.  As we
shall show, contrary to what happens below $d_u$, a {\it different}
Harris criterion establishes when the disorder changes the critical
properties of an infinite system.

In order to investigate whether an infinitesimal disorder affects the
critical properties, let's focus on the critical behavior of a generic
local observable $O_x$, e.g. the average local energy in $x$.  A
simple way to establish the Harris criterion consists in studying the
perturbation induced by the disorder to the critical behavior.  If the
correction, no matter how small it is, ends to be the dominant
contribution close to $T_C$ this means that the disorder is
relevant. Calling $J_y$ the random coupling at site $y$ one obtains
that the corrections due to the disorder are:
\[
\delta O_x= \sum_y \frac{\partial O_x}{\partial J_y} J_y 
\]

This is a random variable whose typical value is given by
\begin{equation}\label{fluct-dis}
\sqrt{\overline{\delta O_x^2}}=\sqrt{\sum_y \left(\frac{\partial O_x}{\partial J_y}\right)^2} \eta
\end{equation}
where $\eta$ is the very small variance of the random couplings, $\overline{J_x J_y}=\eta^2 \delta_{x,y}$.

Below the upper critical dimension, just by scaling or using more
refined techniques \cite{CardybookFSS}, one knows that for the pure
critical systems ${\partial O_x}/{\partial J_y}\propto
1/|x-y|^{d-\alpha/\nu}$ where $\alpha-1$ is the exponent
characterizing the singular part of $O_x$, which is $
|T-T_c|^{-\alpha+1}$.  As a consequence one finds that the disorder
fluctuations scale as $\eta \xi^{-d/2+\alpha/\nu}=\eta
\epsilon^{d\nu/2-\alpha}$. These will become dominant with respect to
the pure critical behavior $ |T-T_c|^{-\alpha+1}$ when $d\nu/2<1$, no
matter how small is $\eta$, if one is close enough to the critical
point.

This is the standard Harris criterion. What does it change above the
upper critical dimension? Actually, the previous derivation can be
repeated identically. The only step where we used that $d<d_u$ is the
assumption on the power law behavior of the response function
${\partial O_x}/{\partial J_y}$. Above $d_u$, one could just use
the mean field critical power law behavior. This already would suggest
that the Harris criterion for the critical properties is different
from the one for FSS.  However, the analysis is tricky because above
$d_u$ one finds that subleading corrections to the critical behavior
dominate the sum in (\ref{fluct-dis}). Let us consider for instance
the $\phi^4$ field theory describing the Ising ferromagnetic
transition and let us take $O_x=\langle \phi_x^2 \rangle$. 
We consider that the random couplings lead to a fluctuating
mass in the field theory, i.e. the disorder couples directly to $O_x$.
In this case the response
function ${\partial O_x}/{\partial J_{x+r}}$ reads below $T_c$ and for $1 \ll r \ll \xi$:
\[
\frac{c}{r^{2d-4}}+(T-T_c)\frac{c'}{r^{d-2}}
\]
where $c$ and
$c'$ are two constants.  Although approaching $T_c$ keeping $r$
finite the first term is the leading one in the above expression, one
finds that the sum in (\ref{fluct-dis}) is dominated by the second
term.  Within mean field theory $\langle \phi_x^2 \rangle$ vanishes linearly with
the temperature at the transition, hence we finds that the disorder
affects the critical properties for $d \nu_u <2$. We verified that
this results holds in more general cases like $\phi^n$ field theories
and with more general bare propagators. It is natural to conjecture
that it holds in general above the upper critical dimension.

In summary, we have found two different Harris criteria. The most
important physical consequence is that in large enough dimension
($d>2/\nu_u$) critical properties of an infinite system will not be
affected by an infinitesimal disorder whereas the FSS maybe affected
even in the infinite dimensional limit depending on the value of the
ratio $d_u\nu_u/2$.

In the following appendix we will give a solvable example of the above
scenario, namely the disordered Blume-Capel model.

\section{\label{sec:A-2} A simple solvable model: the weakly disordered Blume-Capel model.} 

The relevance of sample to sample fluctuations is quite
natural. However, one can be surprised that they affect the dynamical
finite-size scaling and not the thermodynamics. In order to understand
in detail, in a concrete example, the general arguments formulated
above we will consider the following model defined by the Hamiltonian
:
\begin{eqnarray}
H_{BCM}&=&-\sum_{i,j} \xi_{i}\xi_{j} S_{i}S_{j} + \Delta \sum_{i}
S_{i}^{2} - h\sum_{i}\xi_{i} S_{i} \ ,\nonumber \\ & & S_i \in
\{-1,0,1\}
\end{eqnarray}
the $\xi_{i}=1-\delta+\sqrt{\delta}l_i$'s are independent, identically
distributed random variables. $l_i$ are i.i.d. random variables
normalized in such a way that the second and the fourth moment of
$\xi_i$ equals one (this is just a convenient but not at all essential
choice, see below).  When the parameter $\delta=0$ this is the pure
completely connected (or infinite dimensional) Blume-Capel model.  By
increasing $\delta$ one can investigate the role of disorder.  It is
well-known that the pure model, $\delta=0$, has a tri-critical point
characterized by a correlation length exponent $\nu=1/2$ and an upper
dimension $d_u=3$, that is $\nu d_u =3/2$~\cite{BlumeCapelRef}. So
this is indeed a case where the arguments of the previous section
predict that in high dimension the critical properties will not be
affected by disorder contrary to FSS.  Our exact solution will confirm
explicitly this result.

The partition function, for a given disorder realization, is given by:
\begin{eqnarray}
Z &=&\sqrt{\frac{N\beta}{2\pi}} \int dm \ e^{- N\beta f(\beta)}
\nonumber \\ f(\beta)&=&-\frac{1}{2} m^{2} + \frac{1}{\beta N}
\sum_{i} \ln(1+2e^{-\beta\Delta}\cosh\beta(m+h)\xi_{i}) \nonumber
\end{eqnarray}

\begin{figure}[phtb]
\begin{center}
\includegraphics[width=230pt,angle=0]{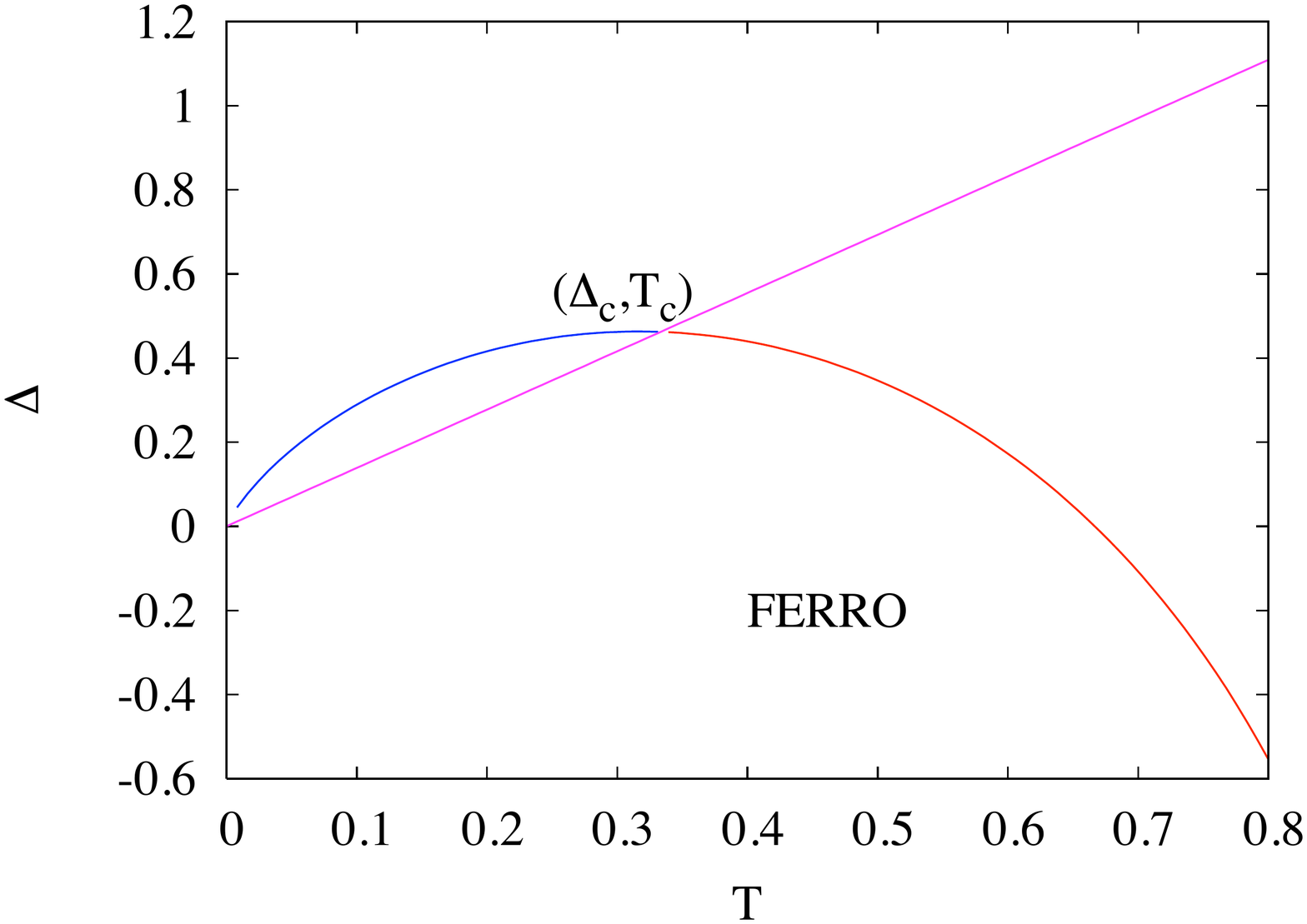}
 \caption{Phase diagram of the pure fully-connected Blume-Capel
   model. }\label{fig:041}
\end{center}
\end{figure}

In order to study the phase transition properties we perform a Landau-like 
expansion of the free energy. One finds at the
sixth order:
\begin{eqnarray}
f(\beta)&=&-\frac{1}{\beta}\ln(1+2e^{-\beta\Delta})+\frac{1}{2}(1-\frac{a(\beta)}{N}\sum_{i}\xi_{i}^{2})\
        m^{2} \nonumber \\ &+&\frac{b(\beta)}{4!N}\sum_{i}\xi_{i}^{4}
        \ m^{4}+\frac{c(\beta)}{6!N} \sum_{i}\xi_{i}^{6} \ m^{6}
        +O(m^{7})
\label{eq:bcm}
\end{eqnarray} 
where $1-a(\beta)$, $b(\beta)$ and $c(\beta)$ are the coefficients of
the Landau expansion of the pure model. They are given by:
\begin{eqnarray}
a(\beta)&=&2\beta\frac{e^{-\beta\Delta}}{1+2e^{-\beta\Delta}}\\
b(\beta)&=&-2\beta^{3}\frac{e^{-\beta\Delta}-4e^{-2\beta\Delta}}{(1+2e^{-\beta\Delta})^{2}}\\
c(\beta)&=&-2\beta^{5}\frac{e^{-\beta\Delta}-26e^{-2\beta\Delta}+64e^{-3\beta\Delta}}{(1+2e^{-\beta\Delta})^{3}} 
\end{eqnarray}
In the pure model, both
$1-a(\beta)$ and $b(\beta)$ may vanish and change of sign. The
cancellation of $1-a(\beta)$ gives the critical temperature, the sign
of $b(\beta)$ the order of the transition. The results for the pure
model are summarized on figure~\ref{fig:041}. The tri-critical point
is defined by the simultaneous cancellation of $1-a(\beta)$ and
$b(\beta)$, that is $(\Delta_{c},T_{c})=(2\ln 2/3,1/3$).
A quick inspection of equation~\ref{eq:bcm} shows that the pure
tri-critical point is not suppressed by disorder. Giving a disorder
realization, $b(\beta)$ vanishes on the line $\beta\Delta=\ln 4$: the
tri-critical point necessarily sits on this line.  More precisely, its
locus is given by:
\begin{eqnarray}
 (\Delta_{c}(N),T_{c}(N))=(\frac{2\ln2}{3N}\sum_{i}\xi_i^2,\frac{1}{3N}\sum_{i}\xi_i^2)
\end{eqnarray}
In the thermodynamic limit this gives back the tri-critical point of
the pure model: $\overline{T_c(N)}=T_c(\infty)\equiv T_c$. Other
choices of normalization of the $l_i$'s would have alter the location
of the tricritical point but not changed the following results on
critical properties and FSS. In the thermodynamic limit (large $N$)
$\frac{1}{N}\sum\xi_{i}^{2}$ is distributed with a Gaussian law of
width $\sigma N^{-1/2}$, where $\sigma$ depends on $\delta$.  This
leads to fluctuations of $T_c$ of order $N^{-1/2}$ and, as a
consequence, similar fluctuations of $\Delta_c$ because of the
relation $\beta\Delta=\ln 4$. This model mimics the ROM case: the
thermal fluctuation given by $\nu d_u$ are of order $N^{-2/3}$ (this
can be checked both for the pure and the disorder model) the disorder
fluctuations, given by the fluctuations of the tri-critical point, of
order $N^{-1/2}$.

To study the tri-critical properties of this model, we focus on the
 line $\beta\Delta=\ln 4$. Writing
 $\beta=\beta_{c}(N)+{\lambda}/{N^{2/3}}$, that is $\lambda
 \propto N^{2/3}(T_{c}(N)-T)$, one obtains at leading order in N:
\begin{eqnarray}
Z(\beta)&=&\sqrt{\frac{N\beta}{2\pi}} \int dm \ e^{ -N\beta f(\beta)}
\nonumber \\
&=&\frac{\alpha\pi^{2}}{\beta_{c}(N)}\sqrt{\frac{\beta}{2}}\left(\frac{3}{2}\right)^{N}N^{1/3}
\mathcal{F}\left(-\frac{\lambda\alpha}{\beta_{c}(N)^{2}}\right) \ ,
\nonumber \\ \mathcal{F}(x)= Ai^2(x)+Bi^2(x) \ &,& \ \alpha=\frac{270}{8} \ 
\label{eq:sca}
\end{eqnarray}
where, $Ai(x)$ and $Bi(x)$ are the well-known Airy functions. They are
two linearly independent solutions of the equation: 
$y^{\prime\prime}-x y =0$.
Their asymptotics, which would be useful to obtain the tails of the
scaling functions of the different thermodynamic observables, is quite
simple:
\begin{eqnarray} 
Ai^{2}(x)+Bi^{2}(x) &\sim&
\frac{1}{\pi\sqrt{-x}} \ \ \ x \leq -1\ , \nonumber\\
Ai^{2}(x)+Bi^{2}(x) &\sim& \frac{e^{\frac{4}{3}x^{3/2}}}{4\pi
x^{1/2}} \ \ \ x\geq 1 \ .
\end{eqnarray}

As a consequence for a given disorder realization, the partition
function has a standard finite-size scaling form like for equation
\ref{eq:fss}, and that $\mathcal{F}$ is independent of the
realization of the disorder.

As sketched in subsection~\ref{sec:4-1}, averaging on the disorder is
here strictly equivalent to average on the distribution of
tri-critical temperatures:
\begin{eqnarray}
p(T_c(N)) &=& \sqrt{\frac{N}{2\pi\sigma^2}} e^{-N\frac{(T_c(N)-T_c)^2}{2\sigma^2}}
\end{eqnarray}

For instance, let us detail the computation of the fluctuations of the
order parameter $m$ defined as:
\begin{eqnarray}
m=\frac{1}{N} \sum \xi_i S_i
\end{eqnarray}
Using the definition of $<m>$, one gets for a given disorder realization:
\begin{eqnarray}
<m>&=& N^{-1/6} \frac{1}{\beta_c(N)}f\left(-\frac{\lambda}{\beta_c(N)^2}\right) +
\mathcal{O} (N^{-5/6}) \nonumber
\end{eqnarray}
with,
\begin{eqnarray*}
f(x)=\frac{51840^{1/3} \pi Bi(\alpha x)+ \frac{45x^2}{2} \
{_{2}F_{1}}(1;4/3,5/3;15x^3)}{\alpha\pi^{3/2}\left(Ai^{2}(x\alpha)+Bi^{2}(x\alpha)\right)}
\end{eqnarray*}
And in a more transparent form:
\begin{eqnarray*}
&(T>T_c(N))& f(x>1) \propto e^{-cst\ x^{3/2}}x^{1/4} \\
&(T<T_c(N))& f(x<-1) \propto (-x)^{1/4}
\end{eqnarray*}
As expected, the tail exponent of the scaling function are those one would have find if one had computed the magnetization by steepest descent. One see clearly, that the magnetization plays in this model the same role as the dynamical overlap.

Similarly, one can compute the non-connected fluctuations of the order parameter,
\begin{eqnarray*}
N<m^2>=N^{2/3}\frac{1}{\beta_c(N)^2} g(-\frac{\lambda}{{\beta_c(N)^2}}) + \mathcal{O}(1)
\end{eqnarray*}
where
\begin{eqnarray*}
g(x)&=&\left((2160)^{1/3}\frac{Ai'(x\alpha)Ai(x\alpha)+Bi'(x\alpha)Bi(x\alpha)}{Ai^2(x\alpha)+Bi^2(x\alpha)}\right)
\end{eqnarray*}
Again the asymptotics of g gives back the thermodynamic exponents. 
\begin{eqnarray*}
&(T>T_c(N))& g(x>1) \propto \frac{1}{x}  \\
&(T<T_c(N))& g(x<-1) \propto \sqrt{x}
\end{eqnarray*}
The non-connected fluctuations have the same behavior as $\chi_4$ for the ROM.
\begin{figure}[phtb]
\begin{center}
\includegraphics[width=280pt,angle=0]{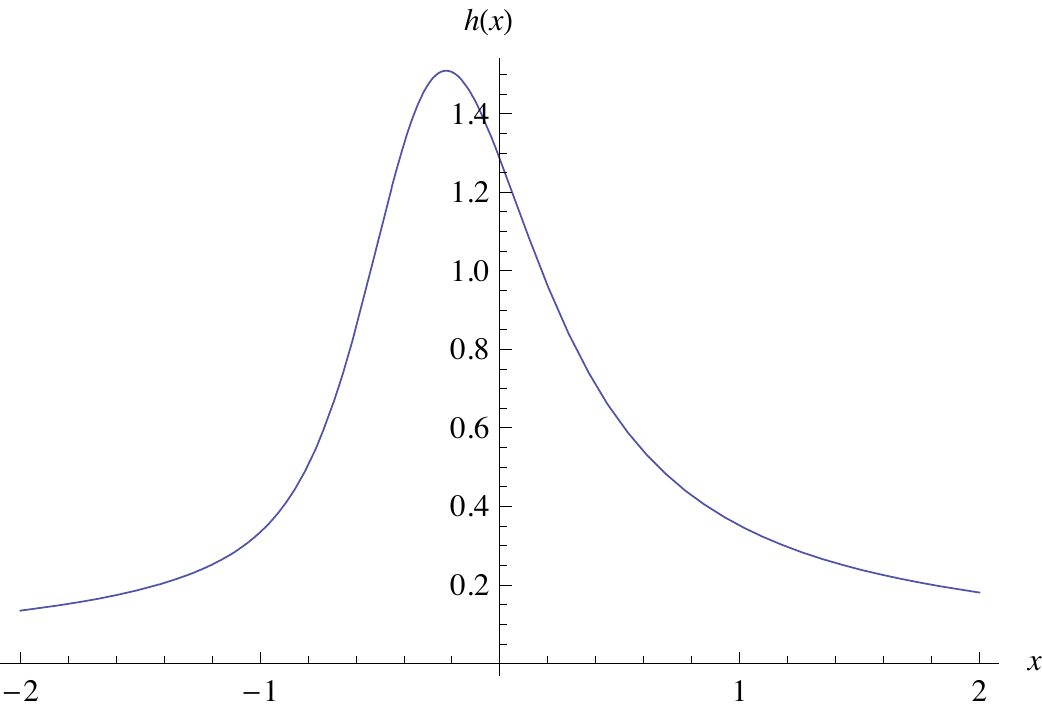}
  \caption{Scaling-function of $N<\delta m ^2>$, on this plot $x>0$
    corresponds to $T>T_c$ }\label{fig:054}
 \end{center}
\end{figure}

To finish this exercise the connected susceptibility is given by, see
figure~\ref{fig:054}:
\begin{eqnarray*}
N<\delta m^2>&=&N^{2/3} h\left(-\frac{\lambda}{{\beta_c(N)^2}}\right) \\
h(|x|>1)&\propto& \frac{1}{x}
\end{eqnarray*}

Now, we have to perform the average over the disorder. For this
purpose, let us introduce $x=\sqrt{N}\frac{T_c(N)-T_c}{\sigma}$. One
has:
\begin{eqnarray}
\overline{\mathcal{O}}(T)&=&\frac{N^{2a/3}}{\sqrt{2\pi}} \int dx
e^{-\frac{x^2}{2}} \mathcal{F}(N^{2/3}(\epsilon)+ N^{1/6}x)
\end{eqnarray}
As showed in the previous appendix, this integral has two parts:
a regular part, corresponding to the small value of $x$, and a
singular part given by the tails of $\mathcal{F}(x)\sim x^{-a}$ for
$x>>1$.

To compute this part, one has to consider all the contributions coming
from the temperature range: $|T-T_c(N)| > c N^{-2/3}$ i.e. such that
$ \sigma x > -N^{1/2}(T-T_c) +cN^{-1/6}$. By defining $\sigma x_0\equiv
N^{1/2}(T-T_c)$, one gets:
\begin{eqnarray}
\overline{\mathcal{O}}(T)&\simeq&\frac{N^{a/2}}{\sqrt{2\pi}} \int_{\sigma(x+x_0) >cN^{-1/6}} \frac{e^{-\frac{x^2}{2}}} {(x_0+ x)^{a}}
\end{eqnarray}

We are now exactly in the same framework as discussed in
section~\ref{sec:4-1}. These simple examples show how the analysis of
this toy model is instructive. It is a fully solvable example of a
weak-disordered model, for $\delta \neq 0$, with an exponent $\nu=1/2$
and an upper dimension $d_u=3$, that is $\nu d_u
=3/2$~\cite{BlumeCapelRef} where FSS is not the standard one. It gives
also a clear insight in understanding the competition of the thermal
and disorder fluctuations.

\section*{References}

\bibliographystyle{iopart-num}

\begin{thebibliography}{10}
\expandafter\ifx\csname url\endcsname\relax
  \def\url#1{{\tt #1}}\fi
\expandafter\ifx\csname urlprefix\endcsname\relax\def\urlprefix{URL }\fi
\providecommand{\eprint}[2][]{\url{#2}}

\bibitem{MCTreviewdas}
Das S~P 2004 {\em Rev. Mod. Phys.\/} {\bf 76} 785

\bibitem{MCTbookGotze}
G\"otze W 2009 {\em Complex Dynamics of Glass Forming Liquids\/} (Oxford
  University Press)

\bibitem{KIR1}
Kirkpatrick T and Wolynes P 1987 {\em Phys. Rev. B\/} {\bf 36} 8552

\bibitem{KIR2}
Kirkpatrick T, Thirumalai D and Wolynes P 1989 {\em Phys. Rev. A\/} {\bf 40}
  1045

\bibitem{KIR3}
Kirkpatrick T and Wolynes P 1987 {\em Phys. Rev. B\/} {\bf 35} 3072

\bibitem{reviewBCKM}
Bouchaud J~P, Cugliandolo F, Kurchan J and M\'ezard M 1996 {\em Physica A\/}
  {\bf 226} 243

\bibitem{Cavagnareview}
Cavagna A 2009 {\em To appear in Phys. Rep.,\/}  cond--mat arXiv:0903.4264

\bibitem{Lubchenko:yq}
Lubchenko V and Wolynes P~G 2007 {\em Annu. Rev. Phys. Chem.\/} {\bf 58} 235

\bibitem{ABB}
Andreanov A, Biroli G and Bouchaud J~P 2009 {\em cond-mat\/}  arXiv:0903.4619

\bibitem{BB-EPL}
Biroli G and Bouchaud J~P 2004 {\em Europhys. Lett.\/} {\bf 67} 21--27

\bibitem{IMCT}
Biroli G, Bouchaud J~P, Miyazaki K and Reichman D~R 2006 {\em Phys. Rev.
  Lett.\/} {\bf 97} 195701

\bibitem{chi4a}
Franz S and Parisi G 2000 {\em J. Phys. Condens. Matter\/} {\bf 12} 6335

\bibitem{BB-SE}
Biroli G and Bouchaud J~P 2007 {\em J. Phys. C\/} {\bf 19} 205101

\bibitem{BBBKKR}
Berthier L, Biroli G, Bouchaud J~P, Kob W, Miyazaki K and Reichman D 2007 {\em
  J. Chem. Phys.\/} {\bf 126} 184503

\bibitem{IndiansPNAS}
Karmakar A, Dasgupta D and Sastry S 2009 {\em PNAS\/} {\bf 106} 3675

\bibitem{Marinari:1992qy}
Marinari E and Parisi G 1992 {\em Europhys. Lett.\/} {\bf 19} 451--459

\bibitem{BIL}
Billoire A, Giomi L and Marinari E 2005 {\em Europhys. Lett.\/} {\bf 71} 824

\bibitem{PAR2}
Marinari E, Parisi G and Ritort F 1994 {\em J. Phys.A: Math.Gen.\/} {\bf 27}
  7647

\bibitem{ParisiPotters}
Parisi G and Potters M 1995 {\em J. Phys. A\/} {\bf 28} 5267

\bibitem{BRA1}
Brangian C, Kob W and Binder K 2002 {\em J.Phys. A: Math Gen\/} {\bf 35} 191

\bibitem{BRA2}
Brangian C, Kob W and Binder K 2002 {\em Phil. Mag. B\/} {\bf 82} 663

\bibitem{BRA3}
Brangian C, Kob W and Binder K 2001 {\em Europhys. Lett.\/} {\bf 53} 756

\bibitem{Brangian:uq}
Brangian C 2003 {\em Physica A\/} {\bf 338} 471--478.

\bibitem{RIT2}
Crisanti A and Ritort F 2002 {\em J.Phys: Condens. Matter\/} {\bf 14} 1381

\bibitem{RIT3}
Crisanti A and Ritort F 2000 {\em Physica A\/} {\bf 280} 155

\bibitem{LEF1}
Cherrier R, Dean D and Lef\`evre A 2003 {\em Phys. Rev. E\/} {\bf 67} 046112

\bibitem{albakrakoviacJCP2002}
Krakoviack V and Alba-Simionesco C 2002 {\em J. Chem. Phys.\/} {\bf 117} 2161

\bibitem{Cugliandolo:ao}
Cugliandolo F 2004 {\em Slow Relaxations and nonequilibrium dynamics in
  condensed matter\/} ({\em Les Houches Summer School\/} vol LXXVII) (Springer
  Berlin / Heidelberg) chap 7: Dynamics of Glassy Systems, pp 367--521

\bibitem{Cavagnapedestrians}
Castellani T and Cavagna A 2005 {\em J. Stat. Mech.\/}  P05012

\bibitem{TAP}
Thouless D, Anderson P and Palmer R 1987 {\em Phil. Mag. A\/} {\bf 35} 593

\bibitem{lalouxkurchan}
Kurchan J and Laloux L 1996 {\em J.Phys. A: Math Gen\/} {\bf 29} 1929--1948

\bibitem{gotzerev}
G\"otze W and Sj\"ogren L 1992 {\em Rep. Prog. Phys.\/} {\bf 55} 241

\bibitem{CHA}
Reichman D and Charbonneau P 2005 {\em J. Stat. Mech.\/}  P05013

\bibitem{gotzehouches}
G\"otze W 1991 {\em Liquids, Freezing and the Glass Transition\/} ({\em Les
  Houches Summer School\/} vol~LI) (North-Holland / Amsterdam) p 287

\bibitem{edigerreview}
Ediger M~D 2000 {\em Annual Review of Physical Chemistry\/} {\bf 51} 99--128

\bibitem{dasguptaetal}
Dasgupta C, Indrani A, Ramaswami S and Phani M 1991 {\em Europhysica Letter\/}
  {\bf 15} 307

\bibitem{parisifranzdonatiglotzer}
Franz S, Donati C, Parisi G and Glotzer S 1999 {\em Phil. Mag. B\/} {\bf 79}
  1827--1831

\bibitem{toninelli:041505}
Toninelli C, Wyart M, Berthier L, Biroli G and Bouchaud J~P 2005 {\em Phys.
  Rev. E\/} {\bf 71} 041505

\bibitem{FranzMontanari}
Franz S and Montanari A 2007 {\em J. Phys. A\/} {\bf 40} F251

\bibitem{Tesi:1996ud}
Tesi M~C, Rensburg E~J~V, Orlandini E and Whittington G 1996 {\em J. Stat.
  Phys\/} {\bf 82} 155

\bibitem{Hukusima:1996yq}
Hukusima K and Nemoto K 1996 {\em J. Phys. Soc. Japan\/} {\bf 65} 1604--1608

\bibitem{Lyubartsev:1992uq}
Lyubartsev A, Martsinovski A and Shevkanov S 1992 {\em J. Chem. Phys.\/} {\bf
  96} 1776--1783

\bibitem{Jacobs:1981uq}
Jacobs L and Rebbi C 1981 {\em Journal of Computational Physics\/} {\bf 41} 203

\bibitem{Krauth:uq}
Santen L and Krauth W 2000 {\em Nature\/} {\bf 405} 550

\bibitem{Billoire:2002fj}
Billoire A and Marinari E 2002 {\em Europhys. Lett.\/} {\bf 60} 775--781

\bibitem{PAR3}
Parisi G and Potters M 1995 {\em J. Phys.A: Math.Gen.\/} {\bf 28} 5267

\bibitem{Bernasconi}
Bernasconi J 1987 {\em J. Phys. France\/} {\bf 48} 559

\bibitem{BM94}
Bouchaud J~P and M\'ezard M 1994 {\em J. Phys. I (France)\/} {\bf 4} 1109

\bibitem{Picco:2000kx}
Picco M, Ritort F and Sales M 2001 {\em The European physical journal. B\/}
  {\bf 19} 565--582

\bibitem{Marinari:1999uq}
Marinari E, Naitza C, Parisi G, Picco M, Ritort F and Zuliani F 1999 {\em Phys.
  Rev. Lett.\/} {\bf 82} 5175

\bibitem{Leutheusser:1984uq}
Leutheusser E 1984 {\em Phys. Rev. A\/} {\bf 29} 2765--2773

\bibitem{Fuchs:1991sf}
Fuchs M, G\"otze W, Hofacker I and Latz A 1991 {\em J. Phys. Condens. Matter\/}
  {\bf 3} 5047

\bibitem{Kim:2001rz}
Kim B and Latz A 2001 {\em Europhys. Lett.\/} {\bf 53} 660

\bibitem{Brezin:1982fj}
Br\'ezin E 1982 {\em J.Phys (Paris)\/} {\bf 43} 15

\bibitem{Brezin:1985kx}
Br\'ezin E and Zinn-Justin J 1985 {\em Nuclear Phys. B\/} {\bf 257} 867--893

\bibitem{Binder:1985kx}
Binder K, Nauenberg M, Privman V and Young A~P 1985 {\em Phys. Rev. B\/} {\bf
  31} 1498--1502

\bibitem{Harris:1974ly}
Harris A 1974 {\em J. Phys. C\/} {\bf 7} 1671

\bibitem{Chayes:1986kx}
Chayes J~T, Chayes L, Fisher D~S and Spencer T 1986 {\em Phys. Rev. Lett.\/}
  {\bf 57} 2999--3002

\bibitem{Aharony:1996ys}
Aharony A and Harris A~B 1996 {\em Phys. Rev. Lett.\/} {\bf 77} 3700--3703

\bibitem{Wiseman:1998vn}
Wiseman S and Domany E 1998 {\em Phys. Rev. E\/} {\bf 58} 2938--2951

\bibitem{GarelMonthus}
Monthus C and Garel T 2007 {\em Markov Processes and Related Fields\/} {\bf 13}
  731--760

\bibitem{Fisher}
Fisher D~S 1995 {\em Phys. Rev. B\/} {\bf 51} 6411

\bibitem{Ioffe:1998df}
Ioffe L~B and Sherrington D 1998 {\em Phys. Rev. B\/} {\bf 57} 7666

\bibitem{Lopatin:2000br}
Lopatin A~V and Ioffe L 2000 {\em Phys. Rev. Lett.\/} {\bf 84} 4208

\bibitem{Lopatin:1999pz}
Lopatin A~V and Ioffe L 1999 {\em Phys. Rev. B\/} {\bf 60} 6412

\bibitem{BBBKKR2}
Berthier L, Biroli G, Bouchaud J~P, Kob W, Miyazaki K and Reichman D 2007 {\em
  J. Chem. Phys.\/} {\bf 126} 184504

\bibitem{LudoFSS}
Berthier L 2003 {\em Phis. Rev. Lett.\/} {\bf 91} 055701

\bibitem{CardybookFSS}
Cardy J~L 1988 {\em Finite-Size Scaling\/} (Elsevier Science Ltd)

\bibitem{Jones:2005uq}
Jones J~L and Young A~P 2005 {\em Phys. Rev. B\/} {\bf 71} 174438

\bibitem{BlumeCapelRef}
Blume M, Emery V~J and Griffiths R~B 1971 {\em Phys. Rev. A\/} {\bf 4}
  1071--1077

\end{thebibliography}

\end{document}